%% file: ebe_review.tex
\documentclass{ws-revolume}  

\usepackage{epsf}

\makeatletter
\def\@dottedtocline#1#2#3#4#5{\ifnum #1>\c@tocdepth \else
  \vskip \z@ \@plus.2\p@
  {\leftskip #2\relax \rightskip \@tocrmarg \parfillskip -\rightskip
    \parindent #2\relax\@afterindenttrue
   \interlinepenalty\@M
   \leavevmode
   \@tempdima #3\relax
   \advance\leftskip \@tempdima \null\hskip -\leftskip
    {#4}\nobreak
        \hfill \nobreak
           \hb@xt@\@pnumwidth{%
             \hfil\normalfont \normalcolor #5}\par}\fi}
\def\numberline#1{\hb@xt@\@tempdima{#1.\hfil}}
\makeatother

\newcommand{\be}{\begin{eqnarray}}
\newcommand{\ee}{\end{eqnarray}}
\newcommand{\non}{\nonumber\\}
\newcommand{\inline}[1]{\noalign{\hbox{#1}}}

\newcommand{\benum}{\begin{enumerate}}
\newcommand{\eenum}{\end{enumerate}}
\newcommand{\bdesc}{\begin{description}}
\newcommand{\edesc}{\end{description}}
\newcommand{\bitem}{\begin{itemize}}
\newcommand{\eitem}{\end{itemize}}
\newcommand{\bhead}{\begin{center}\bf \Large}

\newcommand{\equ}[1]{Eq.~(\ref{#1})}

\newcommand{\Tr}{{\rm Tr}}

\newcommand{\GeV}{\hbox{\,GeV}}
\newcommand{\MeV}{\hbox{\,MeV}}
\newcommand{\fm}{\hbox{\,fm}}

\newcommand{\tw}{\textwidth}
\newcommand{\rb}{\right)}
\newcommand{\lb}{\left(}
\newcommand{\incl}[1]{\overline{ #1}^{\rm incl}}
\newcommand{\reff}[1]{(\ref{#1})}
\newcommand{\ebe}{E-by-E }
\newcommand{\dphase}[2]{\frac{d^#2 #1}{(2 \pi)^#2}}
\newcommand{\QGP}{Quark Gluon Plasma }
\newcommand{\Gave}[1]{\left\langle #1 \right\rangle_{\rm G}}
\newcommand{\Cave}[1]{\left\langle #1 \right\rangle_{\rm C}}
\newcommand{\MBave}[1]{\left\langle #1 \right\rangle_{\rm MB}}
\newcommand{\ave}[1]{\left\langle #1 \right\rangle}
\newcommand{\avesub}[2]{\left\langle #1 \right\rangle_{\rm #2}}
\newcommand{\ebeave}[1]{\left\langle #1 \right\rangle_{\rm ebe}}

\begin{document}

\setcounter{chapter}{0}

\chapter{EVENT BY EVENT FLUCTUATIONS}

\markboth{S. Jeon and V. Koch}{Event by Event Fluctuations}

\author{S. Jeon }
\address{Physics Department, Mc Gill University\\
Montreal, QC H3A-2T8, Canada\\
and \\
Riken-Brookhaven Research Center\\
Brookhaven National Laboratory\\
Upton, NY 11973, USA\\
E-mail: jeon@hep.physics.mcgill.ca}
\author{V. Koch}
\address{Lawrence Berkeley National
Laboratory \\ 
Berkeley, CA, 94720, USA\\
E-mail: vkoch@lbl.gov}

\tableofcontents
\include{introduction}
\include{chapter_1}

\include{chapter_2}
\include{chapter_3}
\include{chapter_4}
\include{chapter_5}

\include{conclusions}
\bibliography{ebe_review}

\end{document}

%% file: introduction.tex
\section{Introduction}

The study and analysis of fluctuations are an essential method to characterize
a physical system. In general, one can distinguish between several classes of
fluctuations. On the most fundamental level there are quantum fluctuations,
which arise if the specific observable  does not commute with the
Hamiltonian of the system under consideration. These fluctuations probably
play less a role for the physics of heavy ion collisions. Second, there are
``dynamical'' fluctuations reflecting the dynamics and responses of the
system. They help to characterize the properties of the bulk (semi-classical)
description of the system. Examples are density fluctuations, which are 
controlled by the compressibility of the system. Finally, there are 
``trivial''
fluctuations induced by the measurement process itself, such as finite number
statistics etc. These need to be understood, controlled and subtracted in
order to access the dynamical fluctuations which tell as about the 
properties of the system.

\begin{figure}[htb]
  \epsfysize=4cm
  \centerline{\epsfbox{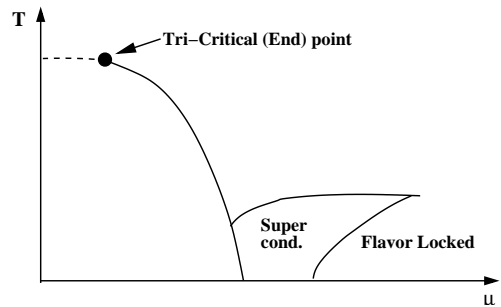}}
  \caption{Schematic picture of the QCD phase-diagram}
  \label{fig:int:phase_diagram}
\end{figure}

Fluctuations are also closely related to phase transitions. The well known
phenomenon of critical opalescence is a result of fluctuations at all
length scales due to a second order phase transition. First order transitions,
on the other hand, give rise to bubble formation, i.e. density fluctuations at
the extreme. Considering the richness of the QCD phase-diagram as sketched in
Fig.\ref{fig:int:phase_diagram}
the study of fluctuations in heavy ions physics should lead to a rich set of
phenomena.

The most
efficient way to address fluctuations of a system created in a heavy ion
collision is via the study of event-by-event (E-by-E) fluctuations, where
a given observable is measured on an event-by-event basis and the fluctuations
are studied over the ensemble  of the events. In most cases (namely when
the fluctuations are Gaussian)
this analysis is equivalent to the
measurement of two particle correlations over the same region of acceptance 
\cite{Bialas:1999tv}.
Consequently, fluctuations tell us about the 2-point functions of the system,
which in turn determine the response of the system to external perturbations.

In the framework of statistical physics, which appears to describe the bulk
properties of heavy ion collisions up to RHIC energies, fluctuations
measure the susceptibilities of the system. These susceptibilities also
determine the response of the system to external forces. For example, by
measuring fluctuations of the net electric charge 
in a given rapidity interval, one
obtains information on how this (sub)system would respond to applying an
external (static) electric field. In
other words, by measuring fluctuations one gains access to the same
fundamental properties of the system as ``table top'' experiments dealing with
macroscopic probes. In the later case, of course, fluctuation measurements
would be impossible.

In addition, the study of fluctuations may reveal information beyond its
thermodynamic properties. If the system expands, fluctuations may be frozen in
early and thus tell us about the properties of the system prior to its
thermal freeze out. A  well known example is the fluctuations in the
cosmic microwave background radiation, as first observed by COBE \cite{Cobe}.

The field of event-by-event fluctuations is relatively new to heavy ion physics
and ideas and approaches are just being developed. So far, most of the
analysis has concentrated on transverse momentum and the net
charge fluctuations.

Transverse momentum fluctuations should be  sensitive to
temperature/energy fluctuations \cite{Stodolsky:1995ds,Shuryak:1998yj}.
These in turn
provide a measure of the heat capacity of the system \cite{Landau_Stat}.
Since the QCD phase transition is associated with a maximum of the specific
heat, the temperature fluctuations should exhibit a minimum in the excitation
function. It has also been
argued \cite{Stephanov:1998dy,Stephanov:1999zu} 
that these fluctuations may provide a signal
for the long range fluctuations associated with the tri-critical point of the
QCD phase diagram. In the vicinity of the critical point the transverse
momentum fluctuations should increase, leading to a maximum of the
fluctuations in the excitation function.

Charge fluctuations \cite{Asakawa:2000wh,Jeon:2000wg},
on the other hand, are sensitive to
the fractional charges carried by the quarks. Therefore, if an equilibrated  
partonic phase has been reached in these collisions, the charge fluctuations
per entropy would be about a factor of 2 to 3 smaller than in an hadronic
scenario.

 In this review, we will systematically examine the principles and the
 practices of fluctuations such as the momentum and the charge
 fluctuations as applied to the heavy ion collisions.   
 In doing so, various concepts of ``average'' need to be introduced. 
 They are: 
 \begin{itemize}
 \item[(i)] Thermal average
 \item[(ii)] Event-by-event average as a whole
 \item[(iii)] Averages over various parts of event-by-event average such as
 clusters and particles emitted by clusters.
 \end{itemize}
 When necessary, we will use subscripts to distinguish these various
 averages.
 However whenever it is clear from the context, 
 or if a relation is of a general nature such as the definition of a
 variance, we will simply use the symbol $\ave{...}$ to denote the average.

 The rest of this review is organized as follows.  
 In section 2, we briefly review the basic concepts of thermal fluctuations
 and its relevance to heavy ion physics.
 In section 3, we examine possible sources of fluctuations other than the
 thermal ones and their effect on experimental measurements.
 In section 4, the relationships between the underlying correlation functions
 and charge fluctuations and also balance functions are established and their
 physical meaning made clear.
 In section 5, various ways of measuring fluctuations proposed so far in
 literature are briefly reviewed and their inter-relationships established.
 In section 6, past and current experimental results are examined in the
 light of the preceding discussions.  
 We conclude in section 7.

 We regret that
 due to the lack of space and also expertise on the authors' side, such
 interesting topics as Disoriented Chiral Condensate or
 Hanbury-Brown-Twiss effects couldn't be discussed in detail
 in this review.  
 Also for the same reason,
 it was impossible for us to cover all the relevant references
 although we tried our best.
 If there is any such glaring omission, we apologies to the authors.


%% file: chapter_1.tex
\section{Fluctuations in a thermal system}
\label{sec:therm_fluct}

To a good approximation the system produced in a high energy heavy ion
collision can be considered to be close to thermal equilibrium. Therefore let
us first review the properties of fluctuations in a thermal system. Most of
this can be found in standard textbooks on statistical physics such as
e.g. Ref.\cite{Landau_Stat} and we will only present the essential points here. 

Typically one considers a thermal system in the grand-canonical
ensemble. This is the most relevant description for heavy ion collisions
since usually only a part of the system -- typically around mid-rapidity -- is
considered. Thus energy and conserved quantum numbers may be exchanged with the
rest of the system, which serves as a heat-bath. There are,
however, important exceptions when the number of conserved quanta is small. In
this case an explicit treatment of these conserved charges is required,
leading to a canonical description of the system
\cite{Cleymans:1991mn} and to significant modifications of the
fluctuations, as we shall discuss below.  
In the following, 
we first discuss fluctuations
based on a grand canonical ensemble and then later point out the differences
to a canonical treatment.

\subsection{Fluctuations in a grand canonical ensemble}
\label{sec:fluct_grand_can}

Assuming we are dealing with a system with one conserved quantum number
(such as the electric charge, baryon number etc.) the grand canonical partition
function is given by\footnote{Here we restrict ourselves to one conserved
  charge. Of course the are several conserved quantum numbers for a heavy ion
  collision. The extension of the formalism to multiple conserved charges is
  straightforward.}
\be
Z = \sum_{\rm states \,\,i} \ave{i \left| 
\exp(-\beta (\hat{H} - \mu \hat{Q})) \right| i}
\equiv  \Tr [ \exp (-\beta (\hat{H} - \mu \hat{Q}) )] 
\label{eq:ch1:part_func}   
\ee
where $\beta = 1/T$ represents the inverse temperature 
of the system, $Q$ is the conserved charge under consideration and $\mu$ is
the corresponding chemical potential.
Here the sum covers a complete
set of (many particle) states. 
The relevant free energy $F$ is related to the partition function via
\be
F = - T \log Z
\label{eq:ch1:free_energy}
\ee
We can further introduce the statistical operator
\be
\hat{\rho}_{\rm G} \equiv \frac{1}{Z} 
\exp \left[ -\beta (\hat{H} - \mu \hat{Q}) \right]
=
\exp \left[ -\beta (\hat{H} - \mu \hat{Q} - F) \right]
\ee
and the moments of the grand-canonical distribution:
\be
\ave{X^n}_{\rm G} \equiv \Tr [ \hat{X}^n\, \hat{\rho}_{\rm G} ] 
\label{eq:ch1:exp_values}
\ee

For a thermal system, typical fluctuations are Gaussian \cite{Landau_Stat} and
are characterized by the variance defined by
\be
\ave{\delta X^2} \equiv \ave{X^2} - \ave{X}^2
\label{eq:ch1:dispersion}
\ee
In the case of grand-canonical ensemble, 
fluctuations of quantities which characterize the thermal
system, such as the energy or the conserved charges, can be expressed in 
terms of appropriate derivatives of the partition function.
Of special interest in the context of heavy ion collisions are
energy/temperature fluctuations, which are often related to the fluctuations
of the transverse momentum as well as electric charge/baryon number
fluctuations. 

\subsubsection{Fluctuations of the energy and of the conserved charges}
\label{sec:dens_fluct}
As already pointed out at the beginning of this section when analyzing the
system created in a heavy ion collisions, one usually studies only a small
subsystem around mid-rapidity. In a statistical framework, this situation is
best represented by a grand canonical ensemble, where the exchange of
conserved quantum number with the rest of the system is taken into
account. The equilibrium state is then characterized by the appropriate 
conjugate variables, namely the temperature and the chemical potentials for
the energy and the conserved quantities, respectively.

As a consequence, energy as well as the conserved charges may fluctuate and
the size of fluctuations reveals additional properties of the matter, the so
called susceptibilities, which also characterize the response of the system to
external forces.

For example, the fluctuation of the conserved charge in the subsystem under
consideration is given by
\be
\Gave{\delta Q^2} = T^2 \frac{\partial^2}{\partial \mu^2} \log Z 
= - T \frac{\partial^2}{\partial \mu^2} F
\label{eq:ch1:charge_fluct_def}
\ee
Similarly, fluctuations of the energy can be expressed in terms of derivatives
of the partition function with respect to the temperature
\be
\Gave{\delta E^2} = \frac{\partial^2}{\partial \beta^2} \log Z = 
- T^3 \frac{\partial^2}{\partial T^2} F = T^2 C_V
\label{eq:ch1:energy_flucts}
\ee
Note that the energy fluctuations are proportional to the heat capacity of the
system, $C_V$. 
Thus one would expect that these fluctuations obtain a maximum as the
system moves through the QCD phase-transition, which, among others, is
characterized by a maximum of the heat capacity.

An alternative way  \cite{Stodolsky:1995ds,Shuryak:1998yj} is to study
temperature fluctuations in {\em micro-canonical} ensemble
which are 
inversely proportional to the heat capacity \cite{Landau_Stat}
\be
\ave{\delta T^2}_{\rm MC} = \frac{T^2}{C_V}.
\label{eq:ch1:temp_fluct}
\ee
However, for situation at hand, which is best described by a grand canonical
ensemble, energy fluctuations appear to be the more appropriate observable.

The second derivatives of the free energy, which characterize the
fluctuations, are usually referred to as susceptibilities. Thus we have the
charge susceptibility
\be
\chi_Q = - \frac{1}{V} \frac{\partial^2}{\partial \mu^2} F
\label{eq:ch1:charge_suscept}
\ee
and the ``energy susceptibility''
\be
\chi_E = - \frac{1}{V} \frac{\partial^2}{\partial T^2} F = \frac{c_v}{T}
\label{eq:ch1:energy_suscept}
\ee
which is related to the specific heat.
And just as the specific heat determines the response of the (sub)system to
a change of temperature the charge susceptibility characterizes the response
to a change of chemical potential. In case of electric charge, this would be
the response to an external electric field. 

In thermal field theory these susceptibilities are given by correlation
functions of the appropriate operators. For example the charge susceptibility
is given by the space-like limit of the static time-time component of the 
electromagnetic current-current
correlator \cite{Kapusta_book,Koch:2001zn,Prakash:2001xm,Doring:2002qa}
\be
\chi_Q = - \Pi_{00}(\omega = 0,q\rightarrow 0)
\label{eq:ch1:chiQ2}
\ee
with
\be
\Pi_{\mu\nu}(k)  
= 
i \int d^4 x \exp(-ikx) \ave{ \hat{\rho} T^* (j_u(x) j_u(0))}
\label{eq:ch1:current_correlator}
\ee

It is interesting to note, that the charge susceptibility is directly
proportional to the electric mass \cite{Kapusta_book}, 
\be
m_{\rm el}^2 =  e^2 \chi_Q  
\ee 
Equation \equ{eq:ch1:chiQ2} allows to calculate the electric mass in any given
model (see e.g. 
\cite{Eletsky:1993hv,Doring:2002qa,Prakash:2001xm,Blaizot:2001vr,Blaizot:2002xz})
and in particular in Lattice QCD 
\cite{Gottlieb:1997ae,Gavai:2002kq}. 
Since dilepton and 
photon production rates are given in terms of the imaginary part of the same
current-current correlation function -- taken at different values of $\omega$
and $q$ -- model calculations for these processes will also give predictions
for the charge susceptibility, which then can be compared with lattice QCD
results. As shown in \cite{Prakash:2001xm} an extraction of $\chi_Q$ from
Dilepton data via dispersion relations, however is not possible. For that one
needs also information for the space-like part of 
$\Pi_{00}$ which is not easily
accessible by  heavy ion experiments.

Charge fluctuations are of particular interest to heavy ion collisions, 
since they provide a signature for the existence of a de-confined Quark Gluon
Plasma phase \cite{Asakawa:2000wh,Jeon:2000wg}. Let us, therefore, discuss
charge fluctuations in more detail. 

Consider a classical ideal gas of positively and negatively charged 
particles of charge $\pm q$. The fluctuations
of the total charge contained in a subsystem of $N$ particles is then given by
\be
\ave{\delta Q^2} 
&=& q^2 \ave{(\delta N_+ - \delta N_-)^2}
\non 
&=& q^2 \left[ \ave{\delta N_+^2} + 
\ave{\delta N_-^2} - 2\ave{\delta N_- \delta N_+}
\right]. 
\ee
Since correlations are absent in an ideal gas, 
$ \ave{\delta N_- \delta N_+} = 0$.
Furthermore, for a classical ideal gas, 
\be
\MBave{\delta N^2} = \MBave{N}
\label{eq:ch1:n_fluct_class}
\ee
where the subscript MB stands for Maxwell-Boltzmann.
This implies
\be
\MBave{\delta Q^2} = q^2 \MBave{N_+ + N_-} = q^2 \MBave{N_{\rm ch}}
\ee
where $N_{\rm ch}= N_+ + N_-$ denotes the total number of charged particles.
Taking quantum statistics into account modifies the results somewhat, since the
number fluctuations are not Poisson anymore (see e.g. \cite{reif_book})
\be
\Gave{\delta N^2} =  \Gave{N}
\left( 1 \pm \Gave{n_\pm} \right)
\equiv \omega_\pm \Gave{N}.
\label{eq:ch1:n_fluct_quantum}
\ee
where 
\be
\Gave{n_\pm} 
\equiv {\int d^3p\, n_\pm(p)^2 \over \int d^3p\, n_\pm(p)} 
\ee
Here, $(+)$  refers to Bosons and $(-)$ to Fermions, and $n_\pm(p)$
represents the respective single particle distribution functions. For the
temperatures and densities reached in heavy ion collisions, however, the 
corrections due to quantum statistics are small. For a pion gas at temperature
$T = 170 \MeV$ $\omega_\pi = 1.13$ \cite{Bertsch:1994qc}.

Obviously, charge fluctuations are sensitive to the {\em square} of the
charges of the particles in the gas. This can be utilized to distinguish a
Quark Gluon Plasma, which contains particles of fractional charge, from a
hadron gas where the particles carry unit charge. 
Charge fluctuations per particle should be smaller in
a Quark Gluon Plasma than in a hadron gas. The appropriate observable to
study is then the charge fluctuations per entropy. 
To illustrate this point, let
us consider a non-interacting pion gas and Quark-Gluon gas in the classical
approximation. Corrections due to quantum statistics and due to the presence of
resonances are discussed in detail in 
\cite{Asakawa:2000wh,Jeon:2000wg,Jeon:1999gr}. In a neutral pion gas the
charge fluctuations are due to the charged pions, which are equally abundant
\be
\MBave{\delta Q_\pi^2} = \MBave{N_{\pi^+}} + \MBave{N_{\pi^-}} 
\label{eq:ch1:dq_pi_gas}
\ee
whereas in a Quark Gluon Plasma the quarks and anti-quarks are responsible for
the charge fluctuations
\be
\MBave{\delta Q^2_{q}} = Q_u^2 \MBave{N_u} + Q_d^2 \MBave{N_d} 
= \frac{5}{9} \MBave{N_q}
\label{eq:ch1:dq_qgp}
\ee
where $N_q = N_u = N_d$ denotes the number of quarks {\em and} anti-quarks. 
For a classical ideal gas of massless particles the entropy is given by
\be
S_{\rm MB} = 4 \MBave{N}
\label{eq:ch1:classical_entropy}
\ee
For a pion gas, this yields
\be
S_\pi = 4 \left( \MBave{N_{\pi^+}} + \MBave{N_{\pi^-}} 
+ \MBave{N_{\pi^0}} \right)
\ee
and for a \QGP 
\be
S_{\rm QGP} &=&
4 \left( \MBave{N_u} + \MBave{N_d} +\MBave{N_{g}} \right)
\non
& = &
4 \left( 2 \MBave{N_q} +\MBave{N_{g}} \right)
\ee
where $N_{g}$ denotes the number of gluons. Therefore, the ratio of charge
fluctuation per entropy in a pion gas is
\be
\frac{\MBave{\delta Q_\pi^2}}{S_\pi} = \frac{1}{6}
\ee
whereas for a 2-flavor  \QGP  it is
\be
\frac{\MBave{\delta Q_q^2}}{S_{\rm QGP}} = \frac{1}{24} 
\ee
Consequently, the charge fluctuations per degree of freedom 
in a \QGP are a factor of four 
smaller than that in a pion gas. Hadronic resonances, which
constitute a considerable fraction of a hadron gas reduce  the result for
the pion gas by about 30 \% \cite{Jeon:2000wg,Jeon:1999gr}, leaving still a
factor 3 signal for the existence of the Quark Gluon Plasma.

The above ratio, $\ave{\delta Q^2}/S$ can also be calculated using
lattice QCD \cite{Gottlieb:1997ae}.
Above the critical temperature, 
the value obtained from lattice
agree rather well with that obtained in our simple \QGP 
model here  \cite{Jeon:2000wg}. More recent calculations for the charge
susceptibility \cite{Gavai:2002kq} give a somewhat smaller value, which would
make the observable even more suitable.

Unfortunately, at present lattice results are not
available for this ratio below the critical temperature. Here one has to
resort to hadronic model calculations. This has been done in
\cite{Doring:2002qa,Eletsky:1993hv} using either a virial expansion, a chiral
low energy expansion or an explicit diagrammatic calculation. In all cases, the
ratio is slightly increased due to interactions, thus enhancing the signal for
the Quark Gluon Plasma.

\subsection{Fluctuations in a canonical ensemble}
\label{sub:fluct_can}
As pointed out in the beginning of this section, once the number of conserved
quanta is small, i.e. of the order of one per event, the grand canonical
treatment, where charges are conserved only on the average, is not adequate
anymore. Instead the description needs to ensure that the quantum number is
conserved explicitly in each event. Since the deposited energy is still large
and is distributed over many degrees of freedom, the canonical ensemble is the
ensemble of choice. 

Obviously the fluctuations of the energy is identical to the grand canonical
ensemble, but fluctuations of particles which carry the conserved charge are
affected. As an example, let us consider the fluctuations of Kaons in low
energy heavy ion collisions. At 1-2 AGeV bombarding energy, only very few
kaons are being produced 
$\ave{N_K} \simeq 0.1$ \cite{Sturm:2002xq}, 
which makes an explicit treatment of strangeness conservation necessary. 
For simplicity, let us consider Kaons and Lambdas/Sigmas\footnote{In the
  following we will denote both Lambdas and Sigmas  as Lambdas, but include
  the appropriate degeneracy factor to take the sigmas into account} 
as the only particles carrying strangeness. 
In the canonical ensemble, where strangeness is
conserved explicitly, the partition function is given 
by\cite{Cleymans:1991mn}
\be
Z_{\rm C} = Z^0_{\rm rest} \sum_{n=0}^{\infty} \frac{1}{n!^2} 
      \left( Z^0_K Z^0_\Lambda \right)^n
\label{eq:ch1:Z_canonic}
\ee
where $Z^0_{\rm rest}$ is the standard (grand canonical) partition for all the
other non strange particles and $ Z^0_K$ , $Z^0_\Lambda$ are the single
particle partition functions for Kaons and Lambdas, respectively.
In the classical limit, these are given by
\be
Z^0_{K} &=& 2 V \int \dphase{p}{3} \exp(-\beta E_{K}) 
= \MBave{N^0_{K}}
\ \\
Z^0_{\Lambda} &=& 4 V \int \dphase{p}{3} \exp(-\beta (E_{\Lambda} -\mu_B))
= \MBave{N^0_{\Lambda}}
\label{eq:ch1:zk0}
\ee
Here, the degeneracy factor of $d=2,4$ for Kaons and Lambdas, respectively, 
take into account the presence of the $K^0$ and $\Sigma$ particles.
Note that $Z^0$ is simply the number of particles in the grand canonical
ensemble in the limit of vanishing strange chemical potential.
Given the above partition function, the probability $P_n$ 
to find $n$ ``Kaons'' is given
by \cite{Ko:2000vp,Jeon:2001ka,Lin:2001mq}
\be
P_n = \frac{\epsilon^n}{I_0(2 \sqrt{\epsilon}) (n !)^2}
\label{eq:ch1:p_n}
\ee
where $I_0$ is the modified Bessel function and 
\be
\epsilon = Z^0_K Z^0_\Lambda = \MBave{N^0_{K}}\MBave{N^0_{\Lambda}}
\ee
Given the probabilities \equ{eq:ch1:p_n}, one can easily calculate  the
canonical ensemble average
\be
\Cave{N_K} &=& \sqrt{\epsilon} 
\frac{I_1(2 \sqrt{\epsilon})}{I_0(2  \sqrt{\epsilon})} 
= \epsilon - \frac{\epsilon^2}{2} + \frac{\epsilon^3}{6} + \ldots 
\non
\Cave{N_K^2} &=& \epsilon
\ee
so that the second factorial moment
\be
F_2^{\rm C} \equiv \frac{\Cave{N(N-1)}}{\Cave{N}^2} 
= \frac{1}{2} + \frac{\epsilon}{6} + \ldots 
= \frac{1}{2} + \frac{\Cave{N_K}}{6} + \ldots
\ee
This is to be contrasted with the grand canonical result, which follows in the
limit of $\epsilon \gg 1$. In this case, 
\be
\Gave{N^2} - \Gave{N} = \Gave{N}^2
\ee
so that
\be
F_{2}^{\rm G} = 1
\ee
Thus for $\ave{N} \ll 1$ the effect of explicit strangeness
conservation reduces the second factorial moment by almost a factor of
two. This suppression of the factorial moment due to explicit charge
conservation can be utilized to measure the degree of equilibration reached in
these collisions (see section \ref{sec:obs_fluct}).

\subsection{Phase Transitions and Fluctuations}
\label{sec:phase_trans}
As already mentioned in the introduction, the QCD phase diagram is expected to
be rich in structure. Besides the well known and much studied transition at
zero chemical potential, which is most likely a cross over transition, a true
first order transition is expected at finite quark number chemical potential. 
It has been argued \cite{Stephanov:1998dy} that the phase separation line
ranges from zero temperature and large chemical potential to finite
temperature and smaller chemical potential where it ends in a critical 
end-point $(E)$ (see Fig.\ref{fig:int:phase_diagram}). Here 
the transition is of second order. 
There is also a first attempt to determine the position of the critical point
in Lattice QCD \cite{Fodor:2001pe}. 

As discussed in
\cite{Stephanov:1998dy}, the associated massless mode carries the quantum
numbers of $\sigma$-meson, i.e. scalar iso-scalar, whereas the pions remain
massive due to explicit chiral symmetry breaking as a result of finite current
quark masses.

Fluctuations are a well known phenomenon in the context of phase
transitions. In particular, second order phase transitions are accompanied by
fluctuations of the order parameter at all length scales, 
leading to phenomena such as critical opalescence \cite{Landau_Stat}. However,
since the system generated in a heavy ion collision expands rather rapidly,
critical slowing down, another phenomenon associated with a second order phase
transition, will prevent the long wavelength modes to fully develop. In
Ref.\cite{Berdnikov:1999ph} these competing effects have been estimated and
authors arrive at a maximum correlation length of about $\xi \simeq 3 \fm$ if
the system passes through the critical (end) point of the QCD phase diagram.

In Ref.\cite{Stephanov:1999zu} the authors argue that if the system freezes out
close to the critical end-point, the long range correlations introduced by the
massless $\sigma$-modes lead to large fluctuations in the pion number at
small transverse momenta. In the thermodynamic limit, this fluctuations would
diverge, but in a realistic scenario, where the long wavelength modes do not
have time to fully develop, the fluctuations are limited by the correlation
length. In Ref.\cite{Berdnikov:1999ph} it is estimated that a correlation length
of $\xi \simeq 3 \fm$ will result in $\sim 5 - 10 \%$ increase in fluctuations
of the mean transverse momentum, which should be observable with present day
large acceptance detectors such as STAR and NA49. Since the precise position of
the critical point is not well known, what is needed is a measurement of the
excitation function of these fluctuations. 

If the system undergoes a first order phase transition, bubble formation may
occur. Since each bubble is expected to decay in many particles this leads to
large multiplicity fluctuations in a given rapidity interval 
\cite{Baym:1999up,Heiselberg:2000ti}. 
Fluctuations of particle ratios, on the other hand, should be reduced due 
to the correlations induced by bubble formation\cite{Jeon:1999gr}.


%% file: chapter_2.tex

\section{Other fluctuations}
\label{sec:other_fluct}

 In heavy ion collisions, there are many sources of
 non-statistical fluctuations.  To extract physically interesting
 information from the observed fluctuations, it is crucial that we know
 the sizes of these other fluctuations.  In this section, we discuss 
 the geometrical volume fluctuations and fluctuations coming purely from the
 initial reactions without further rescatterings within the context of
 wounded nucleon model\cite{Baym:1999up}.

\subsection{Volume fluctuations}
\label{sec:vol_fluct}

 In heavy ion collisions, we have no real control over the impact parameter
 of each collision.  This implies that geometric fluctuation is unavoidable
 in the fluctuations of any extensive quantities such as the particle 
 multiplicity and energy.
 Furthermore, since the system created by a heavy ion collision is finite,
 one must also consider the thermal fluctuation of the reaction
 volume\cite{Landau_Stat}.
 
 To be more specific, consider the multiplicity $N$.  
 Writing
 \be
 N = \rho V
 \ee
 where $\rho$ is the density and $V$ is the volume,
 the fluctuation of $N$ can be written as
 \be
 \ebeave{\delta N^2}
 =
 \ebeave{\delta \rho^2} \ebeave{V}^2 + \ebeave{\rho}^2\ebeave{\delta V}^2 
 \ee
 where the subscript `ebe' indicates the event-by-event average measured in
 an experiment.
 Normally, what we are interested in are the fluctuations in the
 thermodynamic limit.  In other words, we are only interested in
 the fluctuation of the density, $\ave{\delta \rho^2}$.  
 The second term containing $\ave{\delta V^2}$ is
 therefore undesirable.

 To make a rough estimate of the volume fluctuation effect, 
 we first decompose
 \be
 \ebeave{\delta V^2}
 =
 \avesub{\delta V^2}{th}
 +
 \avesub{\delta V^2}{geom}
 \ee
 Assume the ideal gas law $PV = NT$ and
 using
 \be
 \ave{\delta V^2}_{\rm th} 
 =
 -T\left(\partial V\over \partial P\right)_T
 \ee
 (see \cite{Landau_Stat})
 the thermodynamic volume fluctuation can be estimated as
 \be
 \ave{\delta V^2}_{\rm th} = \ave{V}^2/\ave{N}_{\rm th}
 = \ave{V}/\ave{\rho}_{\rm th}
 \ee 
 or
 \be
 \ave{\rho}_{\rm th}^2 \ave{\delta V}^2_{\rm th} = \ave{N}_{\rm th}
 \ee
 which is at least as big as the fluctuation due to the 
 thermal density fluctuation (c.f. Eq.(\ref{eq:ch1:n_fluct_quantum})). 
 
 On top of this, we have purely geometrical fluctuation due to the impact
 parameter variation.
 To have an estimate, let us simplify the nucleus as a
 cylinder with a radius $R$ and the length $L$.
 Assuming geometrically random choice of impact parameters, one can 
 show that
 \be
 {\avesub{\delta V^2}{geom}\over \avesub{V}{geom}^2} =
 {\frac{2\,{{{b_{\rm max}}}^2}}{9\,{{\pi }^2}\,{R^2}}}
 +
 {\frac{16\,{{{b_{\rm max}}}^3}}{27\,{{\pi }^3}\,{R^3}}} 
 +
 O(b_{\rm max}^4/R^4)
 \ee
 if the impact parameter varies between $b=0$ and $b=b_{\rm max}$. The
 expansion is of course valid when $b_{\rm max} \ll R$.

 As an example, consider 6\,\% most central collision of two gold ions.
 For a gold ion, $R\approx 7\,\fm$ and 6\,\% corresponds to 
 $b_{\rm max} \approx 3.0\,\fm$.  Plugging in these values to the above
 formula yields, 
 \be
 \avesub{\delta V^2}{geom} \approx \avesub{V}{geom}^2/170
 \ee
 or
 \be
 \ave{\rho}^2 \avesub{\delta V^2}{geom} = \ave{N}^2/170
 \ee
 For $\ave{N} > 170$ then, the observed
 multiplicity fluctuation can be overwhelmingly due to the impact parameter
 variation.

 This difficulty can be circumvented if the average of the fluctuating
 quantity, say the electric charge $Q$, is zero or close to it.
 This can be readily seen if one writes
 \be
 \delta Q = \delta q\ebeave{V} + \ebeave{q}\delta V
 \ee
 ignoring terms quadratic in $\delta q$ and $\delta V$.
 Squaring and averaging yield
 \be
 \ebeave{\delta Q^2}
 =
 \ebeave{\delta q^2}\ave{V}^2
 +
 \ebeave{\delta V^2} \ebeave{q}^2
 \ee
 assuming that the fluctuations in the charge density $q$ and the volume $V$
 are independent.
 Usually, $\ave{\delta q^2} \sim \ave{N}/\ave{V}^2$ and from the above
 impact parameter consideration, $\ave{\delta V^2} = \ave{V}^2/C$ where
 $C = O(100)$.

 Therefore, the second term is negligible if
 \be
 \ebeave{\delta V^2} \ebeave{q}^2 = \ebeave{Q}^2/C \ll \ebeave{N}
 \ee
 or
 \be
 \ebeave{Q}^2/\ebeave{N} \ll C
 \ee
 The STAR detector at RHIC counts about 600 charged particles per unit 
 rapidity in central collisions and
 $\ebeave{Q}/\ebeave{N} \approx 0.1$.  In that case, 
 \be
 \ebeave{Q}^2/\ebeave{N} \approx (0.1)^2 \times 600 = 6
 \ee
 which is much smaller than $C = O(100)$.
 If one is to talk about gross features, such error would not matter
 much.  However, if a precision measurement is required, this 
 has to be taken into account.

 \subsection{Fluctuation from initial collisions}

 One of the important issues in heavy ion collisions is that of
 thermalization.  Seemingly `thermal' behaviors of particle spectra
 are very common in particle/nucleus collision experiments 
 including $e^+ e^-$ collisions and $pp$, $p\bar p$ collisions where one
 would not expect thermalization among secondary particles to occur.
 This seemingly `thermal' feature results from the way an elementary
 collision distributes the collision energy among its secondary particles. 
 Mathematically speaking, this is analogous to passing from the
 micro-canonical ensemble of free particles to the canonical ensemble of
 free particles.  By doing so, one incurs an error.  However as the size of
 the system grows the error becomes negligible.
 
 A way to distinguish simple equi-partition of energy from true
 thermalization through multiple scatterings is to consider not only the
 average values but also the fluctuations. 
 For instance, as explained in section \ref{sec:therm_fluct}, the
 fluctuations of charged multiplicity in thermal system is Poisson or
 \be
 \MBave{\delta N_{\rm ch}^2} = \MBave{N_{\rm ch}}
 \ee
 However, we do know that in elementary collisions such as
 $pp$ or $p\bar p$, the fluctuation is much stronger due to KNO scaling
 \be
 \ebeave{\delta N_{\rm ch}^2} \propto \ebeave{N_{\rm ch}}^2
 \ee
 
 Questions then arise: 
 How would $\ebeave{\delta N_{\rm ch}^2}$ behave if there is no
 scatterings among the secondaries?  Will it resemble the thermal case or
 retain the features of KNO scaling?  If it turned out to resemble thermal
 case, how can we distinguish?

 To answer these questions, let us invoke the wounded nucleon model as
 explained in Ref.\cite{Baym:1999up}.
 In this model, the final charged particle multiplicity is given by
 \be
 N_{\rm ch} = \sum_{i=1}^{N_p} n_i
 \label{eq:WNM_Nch}
 \ee
 where $N_p$ is the number of participating nucleons (participants) in the
 given event and $n_i$ is the number of charged particles from each of the
 participants.  The assumption here is that the production of charged
 particles from each nucleon is independent and the produced particles do
 not interact further.  
 In that case, it is not hard to show
 \be
 \ebeave{N_{\rm ch}} = \ebeave{N_p}\avesub{n}{N}
 \ee
 and
 \be
 {\ebeave{\delta N^2_{\rm ch}}\over \ebeave{N_{\rm ch}}}
  = 
 {\ebeave{\delta N_p}^2\over \ebeave{N_p}} \avesub{n}{N}
 +
 {\avesub{\delta n^2}{N}\over \avesub{n}{N}}
 \label{eq:dNch_full}
 \ee
 Here $\avesub{...}{N}$ denotes that this average is taken with respect to a
 single nucleon. It is independent of the system size.
 Therefore
 the second term is where KNO makes its appearance. It states that
 that $\avesub{\delta n^2}{N} = c_{\rm KNO} \avesub{n}{N}^2$
 with $c_{\rm KNO} \approx 0.35$ for proton-proton collisions 
 (for instance, see Ref.\cite{Heiselberg:2000fk}).  
 In Ref.\cite{Baym:1995cz}, it is argued that
 since nucleons inside a nucleus are tightly correlated,
 the probability to have all 
 nucleons in the reaction volume participate is very high.
 Therefore $\ebeave{\delta N_p^2}/\ebeave{N_p}$ should be small and 
 $
 {\ave{\delta N^2_{\rm ch}}/\ave{N_{\rm ch}}} = O(\avesub{n}{N})
 $
 which grows with the collision energy. 

 To compare with a typical heavy ion experiment, however, two more effects
 need to be considered in addition.  First, although the fluctuation of
 the number of participants due to the nuclear wavefunction is small, 
 geometrical fluctuations due to the variation of impact parameter can
 introduce more substantial fluctuations in $N_p$.  Second, 
 a heavy ion experiment usually can observe only a portion of the whole
 phase space.
 Therefore when calculating $\ave{n}$ and $\ave{n^2}$
 we must fold in a 
 binary distribution with $p = \ave{n}_{\rm \Delta\eta}/\avesub{n}{full}$
 where $\Delta\eta$ is the detector window.
 This implies replacing
 \be
 & 
 \avesub{n}{N} \to p\ave{n}{N},
 &
 \\
 & 
 \avesub{\delta n^2}{N} \to (1-p)p\avesub{n}{N} + p^2\avesub{\delta n^2}{N},
 &
 \\
 \inline{and}
 &
 \ebeave{\delta N_{p}^2} \to \avesub{\delta N_{p}^2}{geom}
 &
 \ee
 The fluctuation due to the impact parameter variation was estimated in the
 previous section.  Putting everything together then yields
 \be
 \omega_{\rm WNM}
 \equiv
 {\avesub{\delta N^2_{\rm ch}}{ebe}\over \avesub{N_{\rm ch}}{ebe}}
 =
 \avesub{N_{\rm ch}}{geom}
 \left(
 {\frac{2\,{{{b_{\rm max}}}^2}}{9\,{{\pi }^2}\,{R^2}}}
 \right) 
 +
 (1-p) + p\, c_{\rm KNO}\avesub{n}{N}
 \non
 \label{eq:wnm}
 \ee
 Here we assumed that tight correlations renders intrinsic participant
 fluctuation very small and also used the fact that
 $\avesub{N_{\rm ch}}{geom} = p \avesub{N_p}{geom} \avesub{n}{N}$.
 
 A few conclusion can be drawn from Eq.(\ref{eq:wnm}).  First, if 
 $p$ is sufficiently small or if the observational window is sufficiently
 small compared to the whole phase space, the fluctuation of the number of
 charged particle is Poisson.  However, this has nothing to do with
 dynamics.  
 Second, if $p$ is sufficiently large, then 
 $\omega_{\rm WNM}$
 becomes significantly larger than 1 since
 $c_{\rm KNO}\avesub{n}{N}$ is about $7$ at RHIC energy. 
 Third, the ratio $ \omega_{\rm WNM}$ depends linearly on $p$.
 The first point is purely statistical in nature.  The second and third
 points can be used to test the validity of this model where no secondary
 scattering occurs.

 What would change in this consideration if thermalization through multiple
 scatterings occurs? 
 Suppose that the {\em total} number of produced particles is still governed
 by Eq.(\ref{eq:WNM_Nch}).  In other words, only elastic scatterings
 happened to the secondary particles. 
 Further suppose $\avesub{\delta n^2}{N}$
 is still given by the KNO scaling result.  
 In this case, the fluctuations of $N_{\rm ch}$ in the full phase space must
 remain the same as before and Eq.(\ref{eq:dNch_full}) remains valid.  
 Since binary process does not really care about clusters in this case, 
 Eq.(\ref{eq:wnm}) also stays valid.  Therefore elastic scatterings alone
 cannot make any difference in multiplicity fluctuations.

 Now let us relax the condition and allow inelastic collisions among the
 secondaries.  The essential role of inelastic collisions is to convert
 energy into multiplicity and vice versa.  
 Consider a set of events 
 with the same number of participants and same amount of energy deposit.
 According to Eq.(\ref{eq:dNch_full}), the number of initially
 produced particles (we will call them `initial particles') 
 have a distribution that is much wider than the Poisson distribution if 
 $\avesub{n}{N} \gg 1$.
 This wide distribution implies that the 
 distribution of energy per initial particle also has a wide distribution.  
 As argued above, elastic collisions cannot change this situation.
 If inelastic collisions are allowed, then an event with smaller number of
 initial particles would tend to produce more particles 
 since collisions in this event have more available energy.
 In this way, energy is re-distributed evenly and the
 the multiplicity distribution gets narrower.
 This is, of course, the process of equi-partition 
 which lies at the heart of thermalization.   
 Therefore if thermalization does occur starting from a certain energy or a
 certain centrality, $\omega_{N_{\rm ch}}$ must show a corresponding
 behavior changing from approximately $\avesub{n}{N} > 1$ 
 to lower values close to 1 as the energy goes higher or 
 collisions get more central.


%% file: chapter_3.tex

\section{Fluctuations and Correlations}
\label{sec:corr}

Fluctuations measure the width of 
the two particle densities\cite{Bialas:1999tv} and therefore provides
additional information than just the averages.
To illustrate this point, 
let us consider a variable $x(p)$ which depends on momentum of
one particle but does not depend on the multiplicity in each event. 
Let us further consider a generic observable in a given event 
\begin{equation}
S(x) = \sum_{i=1}^N x(p_i) \equiv \sum_{i=1}^N x(i)   
\label{eq:3.1}
\end{equation}
where $N$ is the multiplicity of the  event.  Since $x(p)$ does not depend
on $N$, $S$ is an extensive quantity. 

The {\em event averaged}
moments of this quantity can be expressed as
\begin{equation}
\ebeave{S^k}= \frac1{M} \sum_{m=1}^{M}
\sum_{i_1=1}^{N_m}...\sum_{i_k=1}^{N_m} x_m(i_1)...x_m(i_k)
\label{eq:3.2}
\end{equation}
where $m$ labels the different events and
$M$ is their  total number. $N_m$ is the multiplicity of the event labeled by
$m$.

On the other hand, the moments of an extensive quantity
$x(p)$ calculated from n-particle inclusive
distribution $\rho_n(p_1,...,p_n)$
are defined as
\begin{eqnarray}
\int dp_1...dp_n \rho_n(p_1,...,p_n) [x(p_1)]^{k_1}...
[x(p_n)]^{k_n}  = \nonumber \\  \frac1{M}
\sum_{m=1}^M\sum_{i_1=1}^{N_m}...\sum_{i_n=1}^{N_m}
[x_m(i_1)]^{k_1}...[x_m(i_n)]^{k_n}
\label{eq:3.3}
\end{eqnarray}
where the sums over $i_1...i_n$  include only the terms for which all indices
$i_1...i_n$ are
different from each other.

One sees immediately that (\ref{eq:3.2}) and (\ref{eq:3.3}) are related.
\begin{equation}
\ebeave{S} = \int dp \rho_1(p) x(p)   
\label{3}
\end{equation}
\begin{equation}
\ebeave{S^2} = \int dp_1 dp_2 \rho_2(p_1,p_2) \, x(p_1) x(p_2) 
+ \int d p \rho_1(p) \, [x(p)]^2   
\end{equation}
\begin{eqnarray}
\ebeave{S^3} & = & \int dp_1 dp_2 dp_3 \rho_3(p_1,p_2,p_3) \,x(p_1)x(p_2)x(p_3)
\nonumber \\ 
& & \mbox{}+ 3 \int dp_1dp_2 \rho_2(p_1,p_2) \,x(p_1) [x(p_2)]^2 
\nonumber \\
& & \mbox{} + \int dp \rho_1(p) \, [x(p)]^3    
\label{5}
\end{eqnarray}
Similar formulas can be derived for higher moments of $S$.
By setting $x(p) = 1$, one can also find that the integral over $\rho_n$
gives the $n$-th factorial moments
\be
\int dp_1 ... dp_n\, \rho_n(p_1,...,p_n)
=
\ebeave{N(N-1)\cdots (N-n+1)}
\ee

We have thus established the relation between inclusive measurements and
event-by-event fluctuations
for single particle observables \cite{Gazdzicki:1992ri}, such as e.g. the
total transverse momentum, particle abundances \cite{Gazdzicki:1997gm} etc.

For observables concerning different species of particles,
we define
\be
S_\alpha(x) = \sum_{i=1}^N x(p_i) \delta_{\alpha i}
\ee
where the $\delta$-function picks out the right species from $N$ particles
in the event.
The event average $\ebeave{S_\alpha}$ is just as before with the replacement
of $\rho_1(p) \to \rho_\alpha(p)$ and 
$\int dp \rho_\alpha(p) = \ebeave{N_\alpha}$.
The average number of pairs is then given by
\be
\ebeave{S_\alpha(x) S_\beta(x)}
=
{1\over M}
\sum_{i=1}^{N_m} \sum_{j=1}^{N_m} 
x(p_i) x(p_j)\, \delta_{\beta j} \delta_{\alpha i}
\ee
In terms of the correlation function, this can be rewritten as
\be
\ebeave{S_\alpha(x) S_\beta(x)}
& = &
\int dp_\alpha dp_\beta\, \rho_{\alpha\beta}(p_\alpha, p_\beta) 
x(p_\alpha) x(p_\beta)
\ee
Again by setting $x(p) = 1$, one also obtains 
\be
\int dp_\alpha dp_\beta\, \rho_{\alpha\beta}(p_\alpha, p_\beta) 
=
\ebeave{N_\alpha N_\beta}
\ee

Similar arguments can be constructed for variables which depend on
two or more particle momenta, such as e.g. the fluctuations of 
Hanburry-Brown Twiss (HBT) two particle correlations. 
belong to this class. They are of practical
interest and will be investigated in heavy ion experiments. 
Here we will restrict the argument to two particles but it can be readily
extended to multiparticle correlations. The details are given in 
\cite{Bialas:1999tv}. 

To summarize, \ebe  fluctuations of any (multiparticle)
observable can be re-expressed in terms of inclusive
multiparticle distribution. In case of Gaussian fluctuations the 
multiparticle distributions need to be known up
to twice the order of the observable under consideration.

Following these straightforward considerations, it is natural to discuss
fluctuations in terms of two particle densities, or equivalently
in terms of two
particle correlations functions \cite{Jeon:2001ue,Pruneau:2002yf}

\be
C_{\alpha\beta}(p_1,p_2) 
\equiv \rho_{\alpha\beta}(p_1,p_2) 
- \rho_\alpha(p_2) \rho_\beta(p_1)
\label{eq:3_corr}
\ee
Here the labels $\alpha$ and $\beta$ denote general particle properties
including different quantum numbers. Thus correlations between different
particle types can also be discussed in the same framework.

Instead of the correlation function $C$ often the reduced correlation function
\be
R_{\alpha\beta}(p_1,p_2)
\equiv \frac{C_{\alpha\beta}(p_1,p_2)}{\rho_\alpha(p_1) \rho_\beta(p_2)} = 
\frac{\rho_{\alpha\beta}(p_1,p_2)}{\rho_\alpha(p_1) \rho_\beta(p_2)} - 1
\label{eq:3_red_corr}
\ee
is used. The advantage of using $R_{\alpha\beta}$
over $C_{\alpha\beta}$ is that the trivial scaling with
the square of the number of particles is removed and, therefore, the true
strength of the correlations is exposed more transparently. 
 In particular, as shown in Ref.\cite{Pruneau:2002yf},
 the observable $R_{\alpha\beta}$ has the advantage that it is robust,
i.e. it is independent of the detector efficiencies to leading order.
Also, the analysis of elementary reactions such as proton-proton has
traditionally been done  based on the reduced correlation function
$R_{\alpha\beta}$ 
\cite{Whitmore:1976ip} (see also section \ref{sec:pp-stuff}).

In some cases, such as the charge fluctuations (c.f.~\ref{sec:dens_fluct}),
the fluctuations of the number of particles in a given region of 
momentum space is
considered. In this case one deals with the integrated quantities
\be
\ave{N_\alpha}_{\Delta\eta}
&=& \int_{\Delta \eta} dp_\alpha \rho_1(p_\alpha) 
\\
\ave{N_\alpha N_\beta}_{\Delta\eta}
- \delta_{\alpha\beta} \ave{N_\alpha}_{\Delta\eta}
&=& 
\int_{\Delta \eta} dp_\alpha dp_\beta \rho_2(p_\alpha,p_\beta)
\label{eq:3_N}
\ee
where the notation $\ave{\ldots}_{\Delta\eta}$ is always to be understood as
the event-by-event average in a given momentum space  region $\Delta\eta$
\be
\ave{S}_{\Delta\eta}
=
{1\over M}
\sum_{i=1}^{N_m} \, x(p)\, \theta( p\in \Delta\eta)
\ee
where $\theta(p \in \Delta\eta) = 1$ if $p$ falls inside $\Delta\eta$ and
zero otherwise.
The correlations are then expressed in terms of the `robust
covariances'\cite{Pruneau:2002yf}  
\be
\bar{R}_{\alpha\beta}  
&\equiv& 
\frac{\ave{N_\alpha N_\beta}_{\Delta\eta}
- \ave{N_\alpha}_{\Delta\eta} \delta_{\alpha\beta} 
- 
\ave{N_\alpha}_{\Delta\eta} \ave{N_\beta}_{\Delta\eta} }
{\ave{N_\alpha}_{\Delta\eta} \ave{N_\beta}_{\Delta\eta}} 
\non
& = & \frac{\int_{\Delta \eta} dp_1 dp_2
  \rho_{\alpha\beta}(p_1,p_2) }
  {\int_{\Delta \eta} dp_1 \rho_\alpha(p_1) 
   \int_{\Delta \eta} dp_2 \rho_\beta(p_2)} - 1  
\label{eq:3_rbar}
\ee
which shares the same virtue as $R_{\alpha\beta}$.

\subsection{Correlations and charge fluctuations}
\label{sec:charge_corr}

 We will now concentrate on charge fluctuations to illustrate the above
 general considerations.
 Here we will mainly work out the relationship between the charge
 fluctuations and the charged particle correlation functions. 
 In  full phase space, a conserved charge does not fluctuate, or
 $\ave{\delta Q^2}_{\rm full} = 0$.
 What we are interested in is, however, fluctuations of the net
 charge in a small
 phase space window  where the effect of this overall charge conservation 
 is small. 
 At the same time, this window should be big enough in order not
 to lose information on the widths of the correlation functions.

 From previous section \ref{sec:dens_fluct}, we know that if a QGP forms,
 the charge fluctuation per entropy becomes a factor of 2 to 3 smaller than
 that of the hadronic gas.  What we are interested in this section is how
 that is related to the properties of the charge correlation functions.  
 Since the net charge is $Q = N_+ - N_-$, the variance is
 \be
 \ave{\delta Q^2}_{\Delta\eta}
 =
 \ave{\delta N_+^2}_{\Delta\eta}
 +
 \ave{\delta N_-^2}_{\Delta\eta}
 -2
 \ave{\delta N_+ \,\delta N_-}_{\Delta\eta}
 \ee
 Written this way, it is clear that to have a small charge fluctuation, we
 must have a positive correlation between the unlike-sign pairs and it must
 have enough strength to compensate the independent fluctuations of $N_\pm$. 
 The purpose of this section is to elaborate on these points.

 To write the charge fluctuations using the correlation functions, we first
 decompose
 \be
 \ave{\delta Q^2}_{\Delta\eta}
 & = & 
 \ave{N_+^2}_{\Delta\eta} - \ave{N_+}_{\Delta\eta}^2
 +
 \ave{N_-^2}_{\Delta\eta} - \ave{N_-}_{\Delta\eta}^2
 \non
 & & {}
 -2
 \left[
 \ave{N_+N_-}_{\Delta\eta} - \ave{N_+}_{\Delta\eta}\ave{N_-}_{\Delta\eta}
 \right]
 \ee
 Using Eqs.(\ref{eq:3_corr}) and (\ref{eq:3_N}), we can rewrite this as
 \be
 \ave{\delta Q^2}_{\Delta\eta}
 & = &
 \ave{N_+}_{\Delta\eta}
 +
 \ave{N_-}_{\Delta\eta}
 \non
 & & {}
 +
 \int_{\Delta \eta} dp_\alpha\, dp_\beta\, C_{++}(p_\alpha, p_\beta)
 +
 \int_{\Delta \eta} dp_\alpha\, dp_\beta\, C_{--}(p_\alpha, p_\beta)
 \non
 & & {}
 -
 2 
 \int_{\Delta \eta} dp_\alpha\, dp_\beta\, C_{+-}(p_\alpha, p_\beta)
 \non
 \ee
 The first two terms in the right hand side comes from the fact that
 integrating over like-particle correlations give 
 $\ave{N(N-1)}_{\Delta\eta}$.
 If $\ave{N_+}_{\Delta\eta} \approx \ave{N_-}_{\Delta\eta}$, this can be
 also rewritten in terms of the robust covariances
 \be
 & & 
 {\ave{\delta Q^2}_{\Delta\eta}\over \ave{N_{\rm ch}}_{\Delta\eta}}
 \approx
 1
 +
 {\ave{N_{\rm ch}}_{\Delta\eta}\over 4}\, \nu_{\rm dyn}
 \label{eq:dQ2_R}
 \ee
 where
 \be
 \nu_{\rm dyn}
 \equiv
 \bar{R}_{++}
 +
 \bar{R}_{--}
 -
 2 \bar{R}_{+-}
 \ee
 and $N_{\rm ch} = N_+ + N_-$.
 This is the form advocated in Ref.\cite{Pruneau:2002yf}.
 In this paper, the authors argued that in $\bar{R}_{\alpha\beta}$ 
 the detector efficiencies cancel out while  
 $\ave{N_{\rm ch}}_{\Delta\eta}$ still depends on the efficiency which in
 modern detectors can introduce $10 \sim 20\,\%$ error.

 We know that the single particle distribution $\rho_\pm(p)$ is
 proportional to the 
 probability density function for single particle momentum.  
 To give the two particle distributions similar meaning in terms of
 underlying correlations, 
 let us consider a simple toy model. Consider 
 a gas made up of only three species of resonances, one neutral,
 one positively charged and one negatively charged. No thermal pions are
 present.   
 Furthermore let's further assume that the neutral resonance decays into 
 a pair of
 $\pi^+ \pi^-$ and when a charged resonance decays it emits only one charged
 pions. Since our model does not contain thermal pions, all final state pions
 arise from resonance decay.
 We also assume that
 there is no correlation between the resonances themselves.  
 The probability to have a given set of final state momentum
 $\{\{p_i^+\}, \{p_i^-\}\}$ for charged pions
 is then given by the product of the $\pi^+$ and $\pi^-$ 
 momentum distribution from each resonance decay:
 \be
 \lefteqn{{\cal P}(\{p_i^+\}, \{p_i^-\}|\{P_i\}, M_0, M_+, M_-)}
 \non
 & & {}
 =
 \prod_{a=1}^{M_0} F_0(p_a^+, p_a^-| P_a)
 \prod_{b=1}^{M_+} F_+(p_b^+| P_b)
 \prod_{c=1}^{M_-} F_-(p_c^-| P_c)
 \label{eq:rawP}
 \ee
 where $\{P_i\}$ are the momenta of the resonances just before the decay and
 $M_0, M_+, M_-$ are the number of the neutral and the charged resonances.
 The $F$'s are normalized to 1.
 Since we are not so much interested in the distribution of
 resonance momenta, we integrate
 over it with a suitable weight (for instance thermal weight) before any
 other calculation.  To simplify, we make a further assumption that the
 resonance momenta are uncorrelated.  In that case, 
 \footnote{To be quantum mechanically correct, we need to sum over all
 permutations of $\{p^+\}$ and $\{p^-\}$ and divide by 
 $(M_+ + M_0)! (M_- + M_0)!$. However, since this does not change the
 outcome of our consideration, we will omit that here.} 
 \be
 {\cal P}(\{p_i^+\}, \{p_i^-\}|M_0, M_+, M_-)
 =
 \prod_{a=1}^{M_0} f_0(p_a^+, p_a^-)
 \prod_{b=1}^{M_+} f_+(p_b^+)
 \prod_{c=1}^{M_-} f_-(p_c^-)
 \label{eq:mom}
 \ee
 where 
 \be
 f_0(p_i^+, p_i^-) &=& \int dP\, F_0(p_i^+, p_i^-| P)P_0(P)
 \\
 f_\pm(p_i^\pm) &=& \int dP\, F_\pm(p_i^\pm| P)P_\pm(P)
 \ee
 Here $P_i(P)$ is the momentum distribution for the resonance species $i$.
 
 The function 
 ${\cal P}(\{p_i^+\}, \{p_i^-\}|M_0, M_+, M_-)$
 contains full information about the momentum distribution of the final
 state charged pions given
 the number of underlying resonances. 
 From this distribution all other correlation functions
 can be calculated.  For instance, the single particle distribution for
 positively charged particles is
 \be
 \rho_+(p)
 & = &
 \sum_{M_0, M_+, M_-}
 \int [dp]\, \sum_{i=1}^{M_+ + M_0}\, \delta(p - p_i)\,
 {\cal P}(\{p_i^+\}, \{p_i^-\}|M_0, M_+, M_-)
 \non
 & & {} \qquad\qquad\qquad\qquad\qquad\qquad\times
 {\cal P}(M_0, M_+, M_-)
 \non
 & = &
 \ave{M_0} h_+(p)
 +
 \ave{M_+} f_+(p)
 \label{eq:single_f}
 \ee
 where $[dp]$ is a short hand for integration over all $p$,
 and we defined
 \be
 h_\pm(p_\pm) = \int dp_\mp f_0(p_+, p_-)
 \ee
 Here ${\cal P}(M_0, M_+, M_-)$ is the probability to have 
 the number configuration $(M_0, M_+, M_-)$ in the whole event set and
 $\ave{M_{\pm,0}}$ is the event average of the resonance multiplicities in
 the {\em full} momentum space. Eq.(\ref{eq:single_f})
 thus simply states that
 the number of positively charged  particles is given by the number of
 resonances which decay in positively charged pions time the probability to
 for these pions to have the decay momentum $p$. 

 Likewise, two particle distributions are
 \be
 C_{\pm\pm}(p_1, p_2)
 & = &
%
%
%
 \ave{\delta M_0 \delta M_\pm} h_\pm(p_1) f_\pm(p_2)
 +
 \ave{\delta M_0 \delta M_\pm} h_\pm(p_2) f_\pm(p_1)
 \non & & {}
 +
 \ave{\delta M_\pm^2} f_\pm(p_1) f_\pm(p_2)
 +
 \ave{\delta M_0^2} h_\pm(p_1) h_\pm(p_2)
 \non
 & & {}
 -
 \ave{M_\pm} f_\pm(p_1) f_\pm(p_2)
 -
 \ave{M_0} h_\pm(p_1) h_\pm(p_2)
 \non
 \ee
 and
 \be
 C_{+-}(p_1, p_2)
 & = &
%
%
%
 \ave{\delta M_0 \delta M_-} h_+(p_1) f_-(p_2)
 +
 \ave{\delta M_0 \delta M_+} h_-(p_2) f_+(p_1)
 \non & & {}
 +
 \ave{\delta M_+ \delta M_-} f_+(p_1) f_-(p_2)
 +
 \ave{\delta M_0^2} h_+(p_1) h_-(p_2)
 \non & & {}
 -
 \ave{M_0} h_+(p_1) h_-(p_2)
 +
 \ave{M_0} f_0(p_1, p_2)
\ee
 
 To simplify our consideration,
 let us regard the charged resonances to be iso-spin partners so
 that we have
 \be
 f_+(p) =  f_-(p) \equiv f(p)
 \ee
 For the neutral resonances, $f_0$ should satisfy 
 $f_0(p_1, p_2) = f_0(p_2, p_1)$ which leads to
 \be
 h_+(p) = h_-(p) \equiv h(p)
 \ee
 The average multiplicities are then
 \be
 \ave{N_\pm}_{\Delta\eta} = 
 \ave{M_+}\int_{\Delta\eta} dp\, f(p)
 +
 \ave{M_0}\int_{\Delta\eta} dp\, h(p)
 \ee
 and the charge fluctuation is 
 \be
 \ave{\delta Q^2}_{\Delta\eta}
 & = &
 \ave{N_{\rm ch}}_{\Delta\eta}
 \non
 & & {}
 +
 \left[
 \ave{(\delta M_+ - \delta  M_-)^2}
 -
 \ave{M_+} 
 -
 \ave{M_-} 
 \right]
 \left(
 \int_{\Delta\eta} dp f(p) 
 \right)^2
 \non & & {}
 -2
 \ave{M_0} 
 \int_{\Delta\eta} dp_\alpha\, dp_\beta\, f_0(p_\alpha, p_\beta)
 \label{eq:dQ2DeltaEta}
 \ee

 We can now consider two situations. 
 First, consider the case where the underlying system is a thermal gas of 
 free resonances in the {\em grand canonical ensemble}, 
where charge is only
 conserved on the average. In this case $\MBave{\delta M_i^2} = \MBave{M_i}$ 
 and $\MBave{\delta M_+ \delta M_-} = 0$ (using Boltzmann
 statistics for simplicity) and the above
 reduces to
 \be
 \ave{\delta Q^2}_{\Delta\eta}^{\rm therm.}
 & = &
 \ave{N_{\rm ch}}_{\Delta\eta}
 -2
 \ave{M_0} 
 \int_{\Delta\eta} dp_1\, dp_2\, f_0(p_1, p_2)
 \label{eq:dQ2therm}
 \ee
 If $\ave{\delta Q^2}^{\rm therm.}_{\Delta\eta}$ is to be 
 substantially different from $\ave{N_{\rm ch}}_{\Delta\eta}$, we need to
 have $\ave{M_0} \sim \ave{M_\pm}$ and 
 $ \int_{\Delta\eta} dp_1\, dp_2\, f_0(p_1, p_2)
 \sim 
 \int_{\Delta\eta} dp\, f(p)$.  In other words, the number of neutral
 resonances have to be large and the correlation sharp.
 
 In the full phase space, Eq.(\ref{eq:dQ2therm}) leads to 
 \be
 {\ave{\delta Q^2}_{\rm full}^{\rm therm.}\over 
 \ave{N_{\rm ch}}^{\rm thermal}_{\rm full}}
 & = &
 {\ave{M_+} + \ave{M_-} \over \ave{M_+} + \ave{M_-} + 2\ave{M_0}}
 \label{eq:dQ2full_therm}
 \ee
 which corresponds to the result obtained in Ref.\cite{Jeon:1999gr} for the
 thermal resonance gas.
 To see how the finite phase space result 
 (\ref{eq:dQ2therm}) differs from the result
 (\ref{eq:dQ2full_therm}), consider the extreme case of a flat distribution in
 the pair rapidity and a delta function in the relative rapidity 
 \be
 f(y) &=& {\theta(y_{\rm max} + y)\theta(y_{\rm max} - y) 
 \over 2y_{\rm max}} 
 \label{eq:ext_1}
 \\
 f_0(y_1, y_2)& =& \delta(y_1 - y_2) \, f((y_1 + y_2)/2)
 \label{eq:ext_2}
 \ee
 Here we specified our momentum space variable to be the rapidity $y$ and
 $y_{\rm max}$ is the rapidity of the beam particles in the center of mass
 frame. 
 In this extreme case, it is easy to see that
 \be
 \ave{N_{\rm ch}}^{\rm therm.}_{\Delta\eta} = \ave{M_+ + M_- + 2 M_0} p
 \ee
 and
 \be
 \ave{\delta Q^2}^{\rm therm.}_{\Delta\eta}
 =
 \ave{M_+ + M_-} p 
 \label{eq:therm_ext}
 \ee
 where $p = \Delta\eta/2y_{\rm max}$. Therefore  in this case of infinitely
 sharp correlation,
 \be  
 {\ave{\delta Q^2}^{\rm therm.}_{\Delta\eta}\over
 \ave{N_{\rm ch}}^{\rm therm.}_{\Delta\eta}}
 =
 {\ave{\delta Q^2}_{\rm full}^{\rm therm.}\over
 \ave{N_{\rm ch}}_{\rm full}^{\rm therm.}}
 \label{eq:nice_eq}
 \ee 
 for any values of $\Delta\eta$. 
 
 With more realistic $f$ and $f_0$, 
 the relation (\ref{eq:nice_eq}) does not hold strictly.  
 However as long as 
 \be
 \Delta\eta > \sigma_{y_{\rm rel}} 
 \label{eq:eta_cond}
 \ee
 where $\sigma_{y_{\rm rel}}$ is the width of $f_0$ in the 
 $y_{\rm rel} = y_1 - y_2$
 direction and the
 single particle rapidity distribution is relatively flat within
 $\Delta\eta$,
 Eq.(\ref{eq:nice_eq}) should hold approximately as long as $p$ is not
 too close to 1.
 In this way, one can say that the charge
 fluctuations per charged degree of freedom measured in a restricted 
 rapidity window is a good approximation of the full thermal result.
 From proton-proton collision experiments, we can estimate
 \be
 \sigma_{y_{\rm rel}} \approx 1
 \ee
 Therefore, our rapidity window must be at least that big. 

 As noted previously 
the above results are for the {\em grand-canonical} ensemble.
 In real-life heavy ion collisions, the concept of grand-canonical ensemble
 cannot be applied to the full phase space
 since overall charge conservation strictly requires $\ave{\delta Q^2} = 0$. 
 However, there is no dynamical information in this fact. In particular it
 has nothing to say about whether thermal equilibrium has been established
 within the system. 
 
 To say something about the thermal equilibrium, we need to carve out a
 small enough sub-system and then use the concept of grand-canonical
 ensemble on it.  It would be ideal if we could define a fluid cell within
 the evolving fireball that resulted from the collision of two heavy ions.
 Unfortunately, this is impossible for we have no information on the
 positions of the particles, produced or otherwise.  
 Best we can do is to have (pseudo-)rapidity slices which are supposed to be
 tightly correlated with the spatial coordinates in the beam direction.
 As shown above, if the system {\em is} grand-canonical, 
 then this is not a big problem.
 However, since our underlying system is clearly not, we have to know what
 the effect of having a finite rapidity window is and what the 
 measurements in that restricted window really signify.
 In our opinion, what one should try to get is the right hand-side of 
 Eq.(\ref{eq:dQ2full_therm}) which we know is the right grand-canonical
 limit for the resonance gas.

 To take the finite size effect and charge conservation effect
 into account, we make an restriction 
 $\delta M_+ = \delta M_-$ or
 $M_+ - M_- = Q_c$ is constant.  
 The expression (\ref{eq:dQ2DeltaEta}) now reduces to
 \be
 \ave{\delta Q^2}_{\Delta\eta}
 & = &
 \ave{N_{\rm ch}}_{\Delta\eta}
 -
 \left(\ave{M_+} + \ave{M_-} \right)
 \left(\int_{\Delta\eta} dp\, f(p)\right)^2 
 \non & & {}
 -
 2\ave{M_0} \int_{\Delta\eta} dp_+ dp_-\, f_0(p_+, p_-) 
 \label{eq:main}
 \ee
 which differs from the thermal result 
 (\ref{eq:dQ2therm}) by the second term.
 In the full phase space, this expression results in zero as it should.
 That also means that we need to find a way to 
 extract the
 right-hand-side of Eq.(\ref{eq:dQ2full_therm}) from the above expression. 
 It will be ideal if we can just measure the second term in 
 Eq.(\ref{eq:main}) and subtract it.
 However, to do so we need to know  $\ave{M_\pm}$ which is not readily
 available.

 In the literature, two ways of compensating the charge conservation effect 
have  been proposed.
 A multiplicative correction method was 
 proposed in Ref.\cite{Koch:2001zn} by the present authors
 and an additive correction method was advocated in Ref.\cite{Pruneau:2002yf}.
 To see how the multiplicative corrections work, 
 let us assume for simplicity that $\int_{\Delta\eta}\, f = 
 \int_{\Delta\eta}\, h = p$ and rewrite Eq.(\ref{eq:main}) as
 \be
 \ave{\delta Q^2}_{\Delta\eta}
 & = &
 (1-p) \ave{N_{\rm ch}}_{\Delta\eta}
 -
 2\ave{M_0}
 \left(
 \int_{\Delta\eta} \!\!dp_+ \int_{\Delta\eta} \!\!dp_-\, f_0(p_+, p_-) 
 -
 p^2
 \right)
 \non
 \label{eq:before_corr}
 \ee
 The multiplicatively corrected charge fluctuation is given by 
 \be
 \ave{\delta Q^2}_{\Delta\eta}^{\rm mult.} 
 & = &
 {\ave{\delta Q^2}_{\Delta\eta} \over (1-p)}
 \non
 &= &
 \ave{N_{\rm ch}}_{\Delta\eta}
 -
 2{\ave{M_0}\over (1-p)}
 \left(
 \int_{\Delta\eta} \!\!dp_+ \int_{\Delta\eta} \!\!dp_-\, f_0(p_+, p_-) 
 - p^2
 \right)
 \non
 \label{eq:corrected}
 \ee
 This is the origin of the so called ``$(1-p)$ 
 correction''\cite{Bleicher:2000ek}.
 With the extremely sharp correlation function (\ref{eq:ext_2}) and the flat
 $dN/dy$ (\ref{eq:ext_1}), this yields
 \be
 \ave{\delta Q^2}_{\Delta\eta}^{\rm mult.} 
 &= &
 \ave{N_{\rm ch}}_{\Delta\eta}
 -
 2\ave{M_0}p
 =
 \ave{M_+ + M_-} p
 \ee
 same as Eq.(\ref{eq:therm_ext}).  
 In that case,
 \be
 { \ave{\delta Q^2}_{\Delta\eta}^{\rm mult.} 
 \over \ave{N_{\rm ch}}_{\Delta\eta}^{\rm mult.}}
 =
 { \ave{\delta Q^2}_{\rm full}^{\rm therm.} 
 \over \ave{N_{\rm ch}}_{\rm full}^{\rm therm.}}
 \label{eq:mult_good_eq}
 \ee
 holds for any $\Delta\eta$.
 
 With a more realistic $f$ and $f_0$,  
 Eq.(\ref{eq:mult_good_eq}) does not hold strictly.  
 However, again as long
 the condition (\ref{eq:eta_cond}) holds
 and the single particle rapidity distribution is relatively flat within
 $\Delta\eta$,  
 Eq.(\ref{eq:mult_good_eq}) should hold approximately unless $p$ is very 
 close to 1.

 The authors of Ref.\cite{Pruneau:2002yf} advocated an additive method.
 This method is best explained using the expression (\ref{eq:dQ2_R}).
 Ref.\cite{Pruneau:2002yf} proposes that this expression be corrected by
 replacing  
 $
 \nu_{\rm dyn}
 \to
 \nu_{\rm dyn} + 4/\ave{N_{\rm ch}}_{\rm full} 
 $.  The reason behind this correction is that 
 $-4/\ave{N_{\rm ch}}_{\rm full}$ is the value of $\nu_{\rm dyn}$ when 
 $N_+ - N_-$ is fixed and there is no correlation in the system.
 In our language, this correction corresponds to
 \be
 \ave{\delta Q^2}_{\Delta\eta}^{\rm add.}
 & = &
 \ave{\delta Q^2}_{\Delta\eta}
 +
 p\ave{N_{\rm ch}}_{\Delta\eta}
 \non
 & = &
 \ave{N_{\rm ch}}_{\Delta\eta}
 -
 2\ave{M_0}
 \left(
 \int_{\Delta\eta} \!\!dp_+ \int_{\Delta\eta} \!\!dp_-\, f_0(p_+, p_-) 
 -
 p^2
 \right)
 \label{eq:sergei_corr}
 \ee
 Again with the flat $f(y)$ (\ref{eq:ext_1}) and
 the infinitely sharp $f_0$ (\ref{eq:ext_2}),
 Eq.(\ref{eq:sergei_corr}) yields
 \be
 {\ave{\delta Q^2}_{\Delta\eta}^{\rm add.}\over
 \ave{N_{\rm ch}}_{\Delta\eta}}
 =
 { \ave{\delta Q^2}_{\rm full}^{\rm therm.} 
 \over \ave{N_{\rm ch}}_{\rm full}^{\rm therm.}}
 +
 \left({\Delta\eta\over y_{\rm max}}\right)
 \left(
 { \ave{M_0} \over \ave{M_+ + M_- + 2 M_0}}
 \right)
 \label{eq:sergei_good}
 \ee
 where we used $p = \Delta\eta/2y_{\rm max}$. 
 The second term is negligible only in the $p\to 0$ limit.  
 With a more realistic $f$ and $f_0$, this limit becomes
 \be
 \sigma_{y_{\rm rel}} < \Delta\eta \ll y_{\rm max}
 \ee
 which seems to be of more restrictive use for our purpose of extracting
 the right-hand-side of Eq.(\ref{eq:dQ2full_therm}) (or the first term in
 Eq.(\ref{eq:sergei_good})).
 In the full phase space limit,
 \be
 {\ave{\delta Q^2}_{\rm full}^{\rm add.}\over
 \ave{N_{\rm ch}}_{\rm full}}
 = 
 1
 \ee
 independent of the specific choice for $f_0$. Thus this method
 overcompensates the charge conservation effect slightly.

 One case where we {\em do not} need such corrections is when
 $\ave{M_\pm}=0$ or $\ave{M_\pm} \ll \ave{M_0}$.
 This is the case when all particles are produced in
 neutral clusters.  Exactly this type of situation was analyzed in 
 Ref.\cite{Zaranek:2001di}.  In this reference, however, ``$(1-p)$''
 correction\cite{Bleicher:2000ek} was made which actually overestimated the
 charge fluctuation.  How then do we know the relative strength of 
 $\ave{M_\pm}$ and $\ave{M_0}$?

 From Eq.(\ref{eq:main}), it is clear that the unwanted second 
 term scales like $p^2$.
 The first term $\ave{N_{\rm ch}}$ of course scales like $p$.   
 The last term in Eq.(\ref{eq:main}) however scales differently as the size
 of $\Delta\eta$ varies.  If $\Delta\eta$ is much smaller than the
 correlation length, this term varies like $p^2$.
 However, if 
 $\Delta\eta$ exceeds the correlation length, then integration along 
 $p_{\rm rel} = p_+ - p_-$ ceases to change and hence it varies like $p$.
 This observation suggests the following method:
 Vary the observation window size $\Delta\eta$ and plot 
 $\ave{\delta Q^2}_{\Delta\eta}$ as a function of 
 $p = \ave{N_{\rm ch}}_{\Delta\eta}/\ave{N_{\rm ch}}_{\rm total}$.
 According to Eq.(\ref{eq:main}), this plot can be described by a quadratic
 polynomial in $p$
 \be
 \ave{\delta Q^2}_{\Delta\eta} = a\, p - b\, p^2
 \ee
 If the single particle distribution is flat 
 (as is the case for rapidity distribution), then $p\propto \Delta\eta$
 and  the parameterization 
 $\ave{\delta Q^2}_{\Delta\eta} = a'\, \Delta\eta - b'\, (\Delta\eta)^2$
 can be used as well.

 The coefficients $a$ and $b$ are not truly constants but vary slowly as a
 function of $p$. 
 For small $p$, 
 \be
 a \approx \ave{N_{\rm ch}}_{\rm full} = 
 \ave{M_+ + M_- + 2 M_0}
 \ee
 After $\Delta\eta$ exceeds the correlation length in $f_0$, the 
 $f_0$ term becomes linear in $p$ and the coefficient $b$ becomes. 
 \be
 b \approx \ave{M_+ + M_-}
 \ee
 Therefore by measuring the coefficients in these limits,
 one can estimate the relative strength of $\ave{M_\pm}$ and $\ave{M_0}$ and
 fully correct Eq.(\ref{eq:main}). 

 At this point, one should ask how all this can be reconciled with the {\em
 thermal} fluctuations we considered in previous sections.  
 The results are certainly similar.  However, it seems that we have used
 no thermal features at all in deriving 
 Eqs.(\ref{eq:main}-\ref{eq:corrected}).  
 The expression (\ref{eq:before_corr}) rather suggests that the these
 expressions result from binary process where a particle has a probability
 of $p$ to end up in the detector.  Of course, when $N$ is large and $p$ is
 small in such a way to have $Np$ fixed, binary process becomes Poisson 
 process but that does not mean that the underlying particle energies are
 thermally distributed.  So how can the results of this section 
 be reconciled with the QGP result from previous section?   
 
 Recall that the arguments in section \ref{sec:dens_fluct} that led to our
 conclusion of reduced charge fluctuation depended on two facts.  
 One, the single particle distributions in particular the particle abundances 
are thermal.
 Two, the number fluctuations are all approximately Poisson. 
 As Eq.(\ref{eq:single_f}) shows, the single particle distributions in this
 section do depend on the underlying momentum distribution.  In particular,
 if one is to argue relationships such as 
 Eq.(\ref{eq:ch1:classical_entropy}) hold, one must have thermal momentum
 spectra.  
 On the other hand, it doesn't really matter to the rest of the arguments
 whether the Poisson nature of the number fluctuation 
 (Eqs.(\ref{eq:ch1:dq_pi_gas},\ref{eq:ch1:dq_qgp}) 
 is a result of
 having a thermal system or just a result of having a small detector window.
 Therefore, the fact that Eqs.(\ref{eq:main}, \ref{eq:before_corr}) are not
 the consequences of underlying thermal system does not negate our
 previous conclusion.  
 
 Although the argument given above does reconcile the thermal
 consideration and the results from present section, it is still
 unsatisfactory in some aspects.  The essential point here is our inability
 to have a sub-system in the sense of grand canonical ensemble.
 This is because we can't carve out a sub-system in the coordinate space to
 observe.  
 Unfortunately this limitation is unavoidable. 
 The best we can do is to define sub-systems in (pseudo-)rapidity
 space and argue that strong longitudinal flow strongly correlates the
 (pseudo-)rapidity and the coordinate in the beam direction. 
 It would be desirable to work-out ensemble average taking into account this
 fact.  That analysis, though, is still to be carried out.

 Another question to ask at this point is whether the underlying correlation
 $f_0(p_1, p_2)$ really corresponds to resonances or simply indicates that
 when the pions hadronize from a QGP most of the times they are pair produced.
 One way to distinguish the two
 scenario may be to actually measure the correlation length.  Since the
 unstable neutral meson masses are all larger than twice the pion mass,
 thermal resonance gas
 must exhibit a certain characteristic momentum correlation length between
 the charged pions.  For instance, if a $\rho^0$ at rest decays into two
 pions, they are 3.5 units of rapidity apart along the line of the decay.
 Pions directly coming out of hadronizing QGP on the other hand have no such
 strict constraint imposed on their correlation and should exhibit much
 sharper correlation.  A lack of good hadronization scheme from a QGP makes
 this consideration difficult to substantiate.  However, these points need
 to be further clarified.

\subsection{Correlations and Balance Function}
\label{sec:balance}

 A balance function is a particular way of combining the correlation
 functions to articulate an aspect of the system.  The balance function
 constructed by Pratt et.al.~uses the number of like-sign pairs and
 unlike-sign pairs within a given phase space volume $\Delta\eta$:
 \be
 \lefteqn{B(\eta|\Delta\eta)}
 &&
 \non
 &&
 =
 {1\over 2}\left[
 {\ave{N_{+-}(\eta| \Delta\eta)}\over \ave{N_-(\Delta\eta)}}
 +
 {\ave{N_{-+}(\eta| \Delta\eta)}\over \ave{N_+(\Delta\eta)}}
 -
 {\ave{N_{++}(\eta| \Delta\eta)}\over \ave{N_+(\Delta\eta)}}
 -
 {\ave{N_{--}(\eta| \Delta\eta)}\over \ave{N_-(\Delta\eta)}}
 \right]
 \non
 \ee
 For instance,
 in this expression $N_{+-}(\eta| \Delta\eta)$ is the number of
 unlike-sign pairs which are $\eta$ apart from each other within 
 the window $\Delta\eta$.  To relate $\ave{N_{ij}}$ to 
 to correlation functions, first we express 
 the number of pairs within $\Delta\eta$ and $\eta$ apart as
 \be
 N_{ab}(\eta|\Delta\eta) = 
 \sum_{ij} 
 \theta(p_i^a\in \Delta\eta) 
 \theta(p_j^b\in \Delta\eta) 
 \delta(|p_i^a-p_j^b| - \eta)
 \ee
 Integrating over the expression 
 (\ref{eq:mom}) give 
 \be
 \ave{N_{ab}(\eta|\Delta\eta)}
 & = & 
 \int_{\Delta\eta} dp_1\, dp_2\, 
 \delta(|p_1-p_2|-\eta)\, \rho_{ab}(p_1, p_2) 
 \ee
 With our model of resonance gas, the balance function can be expressed as
 \be
 B(\eta|\Delta\eta)
 & \approx &
 {
 1
 \over
 \ave{N_{\rm ch}}_{\Delta\eta}
 }
 \Big[
 2\ave{M_0} \int_{\Delta\eta} dp_1\, dp_2\, \delta(|p_1-p_2|-\eta)\, 
 f_0(p_1, p_2) 
 \non
 & & {}
 +
 \left(\ave{M_+} + \ave{M_-}\right)
 \int_{\Delta\eta} dp_1 \, dp_2\, 
 \delta(|p_1-p_2|-\eta)\, f(p_1)f(p_2)
 \Big] 
 \non
 \label{eq:balance}
 \ee
 assuming that $\ave{Q}_{\Delta\eta} \ll \ave{N_{\rm ch}}_{\Delta\eta}$.
 From the expressions Eq.(\ref{eq:balance}) and Eq.(\ref{eq:main}) it is
 clear that
 \be
 {\ave{\delta Q^2}_{\Delta\eta}\over\ave{N_{\rm ch}}_{\Delta\eta}}
 \approx
 1 - \int d\eta\, B(\eta|\Delta\eta) 
 \ee
 
 Balance function is originally devised to detect the change in the
 unlike-sign correlation function.  Just as the charge fluctuation, this is
 possible only if $M_0 \sim M_\pm$ and $f_0$ is a sharply peaked function of
 $\eta$.  Pratt et.al.~argued that if a QGP forms, the width of the balance
 function $B(\eta|\Delta\eta)$ will be reduced by a factor of 2 compared to
 the width of the resonance gas balance function.  Recent measurement by
 STAR collaboration\cite{StarBalance}
 indicates that going from peripheral collisions to the
 central collisions there is about 20\,\% reduction in the width of the
 balance function although it is yet not clear whether this is indeed the
 signal of QGP formation.


%% file: chapter_4.tex
\section{Observable fluctuations}
\label{sec:obs_fluct}
Although the motivation for studying fluctuations is similar to that in solid
state physics, the observables in heavy ion collisions are restricted to
correlations in momenta and quantum numbers of the observed particles. 
Spatial correlation are only
indirectly accessible via Fourier transforms of momentum space
correlations, and thus limited. An example is the Bose Einstein correlations
(see e.g. \cite{Heinz:1999rw}). 

Consequently, the fluctuations accessible in heavy ion collisions are all
combinations of (many) particle correlation functions in momentum space. In
addition, as discussed in the previous section, even for tight centrality cuts
there are fluctuations in the impact parameter which may mask the fluctuations
of interest. In the thermal language, these impact parameter fluctuations
correspond to volume fluctuations. Consequently, one should study so called {\em
intensive} variables, i.e. variable which do not scale with the volume, such
as temperature, energy density etc.

Another issue is the presence of statistical fluctuations due to the finite
number of particles observed. These need to be subtracted in order to access
the {\em dynamical} fluctuations of the system.

Finally, although this is outside the expertise of the authors, there are
fluctuations induced by the measurement/detector, which also contribute to 
the signal. Those need to be understood and removed/subtracted as well.

Let us return to the first two issues. As already discussed in some detail in
section \ref{sec:corr} the fluctuations can always be described in terms of
the inclusive many-particle densities in momentum space
$\rho_n(p_1,\ldots,p_n)$. For the simplicity of the discussion let us
concentrate on two particle correlations, which fully characterize
fluctuations of Gaussian nature\footnote{In case of fluctuations of 
positive definite quantities, such as transverse momentum or
  energy,the appropriate distribution is a so called Gamma-
  distribution \cite{Tannenbaum:2001gs}}.

Let us start the discussion by defining 
the number of particles with quantum numbers
$\alpha={\alpha_1,\ldots,\alpha_n}$  in the momentum 
space interval $(p,p+dp)$ in a given {\em event}\footnote{
In this section, we will use the term `event' and `member of the given
ensemble' interchangeably. We also use `event average' and `ensemble average'
interchangeably.
}
\be
n_p^\alpha = \frac{dN_{\rm event}^\alpha}{dp}
\ee 
and its fluctuations 
\be
\delta n_p^\alpha = n_p^\alpha - \ave{n_p^\alpha}
\label{eq:delta_n_p}
\ee
In the case of ideal gas, $\ave{n_p^\alpha}$ takes the form of 
the Bose-Einstein distribution or the Fermi-Dirac distribution depending on
the spin of the particles.

The mean value of an observable\footnote{Here we use sums over momentum
  states, as appropriate for a finite box. The conversion to continuum states
  is a usual, $\sum_p \rightarrow V \int \frac{d^3 p}{(2 \pi)^3}$, where $V$ is
  the volume of the system.}
\be
X &=& \sum_{p,\alpha} x_p^\alpha n_p^\alpha
\label{eq:ch4:X}
\ee
is obtained by averaging over all the events in the ensemble 
\be
\ave{X} &=& \sum_{p,\alpha} x_p^\alpha \ave{n_p^\alpha}
\ee
Note that $X$ defined in this way is an extensive observable.  

The goal of studying fluctuations is to see the effect of dynamics in terms
of non-trivial correlation. 
If the system under study is totally devoid of correlation, then the single
distribution function alone must be able to describe all the moments of $X$
since higher order correlation functions would be
just products of single particle
distribution functions.  Any deviation from this behavior is a sign of
non-trivial correlation.  
To make a quantitative calculation, let us define the single particle {\em
probability} density function
\be
\incl{n_p^\alpha} = {1\over \ave{N_\alpha}} \rho_\alpha(p)
\label{eq:ch4:np_incl}
\ee 
where $\rho_\alpha(p)$ is the single particle inclusive distribution
function introduced in section \ref{sec:corr}.
Note the factor $1/\ave{N_\alpha}$
in the definition of $\incl{n_p^\alpha}$ which 
normalizes this distribution to unity.  Note also that event average has
already performed and the only variable left to average over is the single 
momentum.  We define the average taken with this probability 
density as the `inclusive average'.
\be
\incl{X} = \sum_{p,\alpha} x_p^\alpha \incl{n_p^\alpha}
\ee
For a single particle observable $X$
\be
\incl{X} = \frac{\ave{X}}{\ave{N}} 
\ee   


For the discussion of fluctuations the relevant quantity to
consider is
\be
\Delta_{p,q}^{\alpha,\beta} \equiv 
\ave{\delta n_p^\alpha \delta n_q^\beta} 
\label{eq:ch4:basic_corr}
\ee
where $\delta n_p^\alpha$ is defined in Eq.(\ref{eq:delta_n_p}). 
For an ideal gas, only the identical particles are correlated 
and $\ave{(\delta n_p^\alpha)^2} = \omega_p^{\alpha} \ave{n_p^\alpha}$.
Therefore,
\be
\Delta_{p,q}^{\alpha,\beta} 
= \delta_{pq} \lb  \prod_i \delta_{\alpha_i\beta_i} \rb \, \omega_p^\alpha \,
\ave{n_p^\alpha}. 
\label{eq:ch4:delta_ideal}
\ee
Here 
\be
\omega_p^\alpha = 1
\ee 
for a classical ideal gas, and 
\be
\omega_p^\alpha = (1 \pm \ave{n_p^\alpha})
\label{eq:ch4:omega_ideal} 
\ee
for a boson ($+$) and a fermion ($-$) gas.  Before we continue, let us step
back and try to understand eqs. \reff{eq:ch4:delta_ideal} to
\reff{eq:ch4:omega_ideal} in simple terms. The term $\prod_i
\delta_{\alpha_i\beta_i}$ ensures that only identical particles are
correlated. For example, $\pi^+$ and $\pi^-$, are both pions but with different
charge. Therefore they are not correlated in an ideal gas, i.e. 
$\ave{\delta  \pi^+ \delta \pi^-} =0$. Also there are no effect due to quantum
statistics, as the particles are different.

In the presence of dynamical correlations, the basic correlator
\reff{eq:ch4:basic_corr} may contain off-diagonal elements 
in momentum space and/or in the space of the quantum numbers. For
example the presence of a resonance, such as a $\rho_0$,  
correlates the number of $\pi^+$ and $\pi^-$ in the final state due to
resonance decay. Also their momenta are correlated. In addition, 
the occupation number $n_p$ in a 
given momentum interval may be changed. This would be for instance an effect
of hydrodynamical flow.

Any two-particle observable can be expressed in terms of the basic correlator
$\Delta_{p,q}^{\alpha,\beta}$ \cite{Bialas:1999tv,Stephanov:1999zu}. 
For the generic observable $X$ as given by eq. \reff{eq:ch4:X} the variance 
is given by
\be
\ave{\delta X^2} &=& \sum_{p,q,\alpha,\beta} \Delta_{p,q}^{\alpha,\beta} 
x_p^\alpha x_q^\beta
\ee

To illustrate the formalism, let us consider the net charge\footnote{Note 
the difference in notation: whereas 
$\delta n_p$ refers to the fluctuations in the momentum interval 
$(p,p+dp)$, $\delta N$ refers 
to the fluctuations of the total (integrated) number of particles.}
\be
Q = N_+ - N_- = \sum_{p,\alpha} q_\alpha\, n_p^\alpha 
\label{eq:ch4:Q}
\ee
Here, $q_\alpha$ is the charge of the particle whereas  
all other quantum numbers are still being summed over. Likewise the number of
charged particle is given by
\be
N_{\rm ch}= N_+ + N_- = \sum_{p,\alpha} \, |q_\alpha|\, n_p^\alpha 
\label{eq:ch4:N_ch}
\ee
The variances of these quantities are given by
\be
\ave{ \delta Q^2} &=& \ave{\delta N_+^2} +\ave{\delta N_-^2} - 2
\ave{\delta N_+ \delta N_-} 
\non
&=& \sum_{p,q} \lb \Delta_{p,q}^{+,+} +
\Delta_{p,q}^{-,-} - 2 \Delta_{p,q}^{+,-} \rb
\label{eq:ch4:delta_Q}
\\
\ave{ \delta N_{\rm ch}^2} &=& \ave{\delta N_+^2} +\ave{\delta N_-^2} + 2
\ave{\delta N_+ \delta N_-} 
\non
&=& \sum_{p,q} \lb \Delta_{p,q}^{+,+} +
\Delta_{p,q}^{-,-} + 2 \Delta_{p,q}^{+,-} \rb
\label{eq:ch4:delta_n_ch}
\ee
Here the basic correlator $\Delta_{p,q}^{+,+}$ still contains a sum over all
other quantum numbers, which we have suppressed. 
\be
\Delta_{p,q}^{+,+} = \sum_{\alpha_i \ne {\rm charge}, \beta_i \ne {\rm
    charge}}\Delta_{p,q}^{\{+,\alpha_i\},\{+,\beta_i\}}
\ee 
and similar for the other combinations. This sum simply means that all charged
particles are included independent of their flavor, spin etc.
In the following, we will always use
this short-hand notation with implicit summation over all the quantum numbers
not specified.

Obviously all the information revealed by the fluctuations 
is encoded in the basic correlator $\Delta_{p,q}^{\alpha,\beta}$ and the 
fluctuations of different observables expose different ``moments'' and
elements of the  basic correlator. 
Using the notation from the previous section, the basic
correlator for $\alpha\ne\beta$
is nothing but the 2-particle correlation function defined in 
Eq.(\ref{eq:3_corr}).
For the identical particle correlator, 
Eq.(\ref{eq:ch4:basic_corr}) differs from Eq.(\ref{eq:3_corr}) by a single
particle distribution so that 
\be
\Delta_{p,q}^{\alpha,\beta}
& = & 
\rho_{\alpha\beta}(p,q) - \rho_\alpha(p) \rho_\beta(q) 
+ 
\delta_{\alpha\beta}\,\delta_{pq}\, \rho_\alpha(p) 
\non
& = &
C_{\alpha\beta}(p,q)
+ 
\delta_{\alpha\beta}\,\delta_{pq}\, \rho_\alpha(p) 
\ee

Therefore, an explicit measurement of the two-particle correlation function 
$C_{\alpha\beta}$ would be extremely valuable,
since all (Gaussian) fluctuation
information can be extracted from it by properly weighted integrals.

If the system has no genuine two particle correlations, then the basic
correlator will be
\be
\Delta_{p,q}^{\alpha,\beta} = \ave{n_p} \delta_{p,q} \delta_{\alpha,\beta}
\label{eq:ch4:no_corr}
\ee
similar to the ideal gas, only that $\ave{n_p}$  
does not have to follow a Boltzmann
distribution. Using the inclusive single particle spectrum $\incl{n_p}$
\reff{eq:ch4:np_incl}  as momentum distribution, the relation
\reff{eq:ch4:no_corr} may be utilized as an estimator of the fluctuations due
to finite number statistics\cite{Gazdzicki:1992ri}, which follow Poisson
statistics. A more detailed discussion on this aspect will be given below.

Any correlations, on the other hand will lead to a non Poisson
behavior of the basic correlator. The ideal quantum gases discussed above
are examples where the symmetry of the wavefunction introduces 2-particle
correlations which either reduce (Fermions) or enhance (Bosons) the
fluctuations. Also global conservation laws such as charge, energy etc. will
affect the fluctuations. However, if only a small subset of the system is
discussed these constraints will be negligible\footnote{This is the same
  argument which leads to the canonical and grand-canonical description of a
  thermal system, as briefly discussed in section \ref{sec:therm_fluct}.}. As
the fraction of accepted particles increases, these constraints may become more
relevant, however, and thus need to be properly accounted for. This has been
discussed in some detail in the previous section.

As mentioned in the beginning of this section, in order to avoid contributions
from volume / impact parameter fluctuations it is desirable to study so
called intensive quantities, i.e. quantities which do not scale with the size
of the system. Examples which are currently explored experimentally are the
mean transverse momentum and mean energy fluctuations as well as fluctuations
of particle ratios. 

\subsection{Fluctuations of Ratios}
Let us begin with a general discussion of fluctuations of ratios, since mean
energy or mean transverse momentum as well as particle ratios are all ratios
of observables. Consider the ratio of of two observables $A$ and 
$B$\footnote{Defined this way, both $A$
and $B$ scale with $N$, the number of particles in the final state.} 
\be
A = \sum_{p,\alpha} a_p^\alpha n_p^\alpha 
\non
B = \sum_{p,\alpha} b_p^\alpha n_p^\alpha.
\ee
To find the average and the fluctuation of the ratio
\be
R_{AB} \equiv \frac{A}{B}
\ee
we first write $A = \ave{A} + \delta A$ and $B = \ave{B} + \delta B$
and expand the numerator and the denominator to get
\be
R_{AB}
=
{\ave{A}\over \ave{B}}
+
\frac{\ave{A}}{\ave{B}} \left( \frac{\delta A}{\ave{A}} - 
\frac{\delta B}{\ave{B}} \right)
+
O(\delta^2)
\label{eq:ratio_exp}
\ee
where $O(\delta^2)$ indicates terms that are of quadratic and higher orders
in $\delta A$ and $\delta B$.  Since $A$ and $B$ 
are extensive observables, the
neglected terms are at most ${\cal O}(1/\ave{N})$ which we will neglect from
now on.
From Eq.(\ref{eq:ratio_exp}), it is easy to see that
\be
\ave{R_{AB}} = \frac{\ave{A}}{\ave{B}}
\ee
and the variance
\be
\ave{\delta R_{AB}^2} 
= \frac{\ave{A}^2}{\ave{B}^2} \lb 
\frac{ \ave{\delta A^2}}{\ave{A}^2} +  \frac{ \ave{\delta B^2}}{\ave{B}^2} 
- 2 \frac{\ave{\delta A \,\delta B}}{\ave{A}\ave{B}} \rb
\label{eq:ch4:delta_r1}
\ee

Using the basic correlator $\Delta_{p,q}^{\alpha,\beta}$ 
this can be rewritten as
\be
\ave{\delta R_{AB}^2} = \frac{\ave{A}^2}{\ave{B}^2} 
\sum_{p,q} \sum_{\alpha,\beta}  \Delta_{p,q}^{\alpha,\beta}
\lb 
\frac{a_p^\alpha a_q^\beta}{\ave{A}^2} + 
\frac{b_p^\alpha b_q^\beta}{\ave{B}^2}  
-2 \frac{a_p^\alpha b_q^\beta}{\ave{A} \ave{B}} 
\rb
\label{eq:ch4:ratio_fluct}
\ee
For an ideal gas, this simplifies to
\be
\ave{\delta R_{AB}^2}^{\rm ideal}
= \frac{\ave{A}^2}{\ave{B}^2} \sum_{p} \sum_{\alpha} 
\ave{n_p^\alpha} \omega_p^\alpha
\lb \frac{(a_p^\alpha)^2}{\ave{A}^2} + 
\frac{(b_p^\alpha)^2}{\ave{B}^2}  
-2 \frac{a_p^\alpha b_p^\alpha}{\ave{A} \ave{B}} \rb
\ee
Obviously, deviations from the ideal gas value will give us insight into the
dynamical correlations of the system. In order to expose these deviations,
very often one compares to the fluctuations based on the `inclusive' single
particle distribution. The latter estimates the contribution of finite number
fluctuations to the observed signal. In our formalism, this means that one
evaluates equation \reff{eq:ch4:ratio_fluct} with the fully uncorrelated basic
`correlator' as defined in equ. \reff{eq:ch4:no_corr}. In addition, 
as it is common in the literature, one replaces the event averages
$\ave{\ldots}$ by inclusive averages $\incl{\ldots}$, which simply mean
multiplying by the appropriate factors of $\ave{N}$. This let us define
\be
\lefteqn{
\incl{(\delta R)^2}  \equiv  
}
\non
&&
\frac{(\incl{A})^2}{(\incl{B})^2} \sum_{p} \sum_{\alpha} 
\incl{n_p^\alpha}
\lb \frac{(a_p^\alpha)^2}{(\incl{A})^2} + 
\frac{(b_p^\alpha)^2}{(\incl{B})^2}  
-2 \frac{a_p^\alpha b_p^\alpha}{\incl{A} \incl{B}} \rb.
\label{eq:ch4:estimator}
\ee 
In absence of any dynamical correlations
\be
\ave{(\delta R)^2} = \frac{1}{\ave{N}} \incl{(\delta R)^2}.
\ee
This observation has lead to the measure  of dynamical fluctuations 
$\sigma_{\rm dynamic}^2$ \cite{Voloshin:1999a}
which is given by 
\be
\sigma_{\rm dynamic}^2 = \ave{(\delta R)^2} - \frac{1}{\ave{N}} 
\incl{(\delta R)^2}.
\label{eq:ch4:sig_dyn}
\ee
The first such measure to be proposed has been the so called $\Phi$
variable \cite{Gazdzicki:1992ri,Mrowczynski:1998vt} which in terms of our variables
here is given by
\be
\Phi \equiv \sqrt{\ave{N} \ave{(\delta R)^2}} -
\sqrt{ \incl{(\delta R)^2}}.
\ee
The  ratio has also been proposed  \cite{Stephanov:1999zu} 
\be
F \equiv \frac{ \ave{N} \ave{(\delta R)^2}}{\incl{(\delta R)^2}}
\ee
which has the advantage of being dimensionless.

Aside from simply subtracting the expected fluctuations from an uncorrelated
system, the sub-event method has been developed in \cite{Voloshin:1999a}.

We note, that given the same momentum distribution 
the inclusive result is up to a trivial factor  
identical to that of an classical ideal
gas, i.e $\sigma_{\rm dynamic} = \Phi = 0$ or $F =1$ in this case. 
However, in reality, the inclusive momentum
distribution usually differs from a Boltzmann shape due to additional effects
such as hydrodynamic flow. Notice also, that the 
results for Bose and Fermi gases
already differ from the inclusive estimator, reflecting the correlations
induced by quantum statistics. For a Bose gas, $\Phi > 0$, $F > 1$, whereas
for a Fermi gas $\Phi < 0$, $F < 1$. For the observable to be discussed below, 
the corrections to $F$ are a few percent \cite{Stephanov:1999zu}. 

Finally, the above formalism allows us to discuss more general correlations
between ratios of observables. Lets introduce two more observables $C$ and $D$
\be
C = \sum_{p,\alpha} c_p^\alpha n_p^\alpha, \,\,\,\,
D = \sum_{p,\alpha} d_p^\alpha n_p^\alpha.
\ee
Then, the most general correlation of ratios can be written as 
\be
\lefteqn{
\ave{\delta \lb \frac{A}{B} \rb  \delta \lb \frac{C}{D} \rb } =
}
\non
&&
\frac{\ave{A}\ave{C}}{\ave{B}\ave{D}} 
\lb 
\frac{\ave{\delta A \delta C}}{\ave{A}\ave{C}} 
+\frac{\ave{\delta B \delta D}}{\ave{B}\ave{D}}
-\frac{\ave{\delta A \delta D}}{\ave{A}\ave{D}} 
-\frac{\ave{\delta B \delta C}}{\ave{B}\ave{C}}
\rb
\ee
Using the basic correlator \reff{eq:ch4:basic_corr} we obtain
\be
\lefteqn{
\ave{\delta \lb \frac{A}{B} \rb  \delta \lb \frac{C}{D} \rb } =
\frac{\ave{A}\ave{C}}{\ave{B}\ave{D}} 
}
\non
&&
\sum_{p,q,\alpha,\beta} 
\lb 
\frac{a_p^\alpha c_q^\beta}{\ave{A}\ave{C}} 
+\frac{b_p^\alpha d_q^\beta}{\ave{B}\ave{D}}
-\frac{a_p^\alpha d_q^\beta}{\ave{A}\ave{D}} 
-\frac{b_p^\alpha d_q^\beta}{\ave{B}\ave{C}}
\rb
\Delta_{p,q}^{\alpha,\beta}
\label{eq:unlike_corr}
\ee
Note, that again this correlation function represents nothing but a specific
moment of the basic correlator. Obviously, the replacements $C = A$ and  $D=B$
leads to the ratio fluctuations discussed above. This general correlation will
become useful below, when discussing charge dependent and charge independent
transverse momentum fluctuations, as proposed by the STAR collaboration
\cite{Ray:2002md}. 

After having developed the general formalism for fluctuations of ratios it is
now straight forward to discuss the actual observables.

\subsection{Fluctuations of the mean energy and mean transverse momentum}
The mean energy and transverse momentum are defined as
\be
\epsilon &=& \frac{E}{N} = \frac{\sum_{p} n_p E(p)}{\sum_p n_p} 
\non
p_t      &=& \frac{P_t}{N} = \frac{\sum_{p} n_p p_t(p)}{\sum_p n_p} 
\ee
where $E$ and $(P_{t})$ denote the Energy and transverse momentum of
an event with $N$ particles in the final state. Obviously, these
observables are ratios and the above formalism can be readily applied with the
substitutions
\be
A = E; \,\,\, &{\rm or}& \,\,\, A = P_t
\non
B & = & N
\ee
So that for $X$ being either the transverse momentum of the energy 
\be
\ave{\lb \delta \frac{X}{N} \rb^2} 
&=& \frac{1}{\ave{N}^2} 
\ave{\left( \delta X -  \frac{\ave{X}}{\ave{N}} \, \delta N \right)^2} 
\ee
After rearranging terms one finds for an ideal gas \cite{Stephanov:1999zu}
\be
\ave{\lb \delta \frac{X}{N} \rb^2} = \frac{1}{\ave{N}^2} \sum_p \left[ 
\lb x_p - \ave{\frac{X}{N} } \rb^2 \omega_p \ave{n_p}\right]
\ee
where $X$ stands either for the energy $E$ or the transverse momentum $P_t$.

In addition to the transverse momentum fluctuations for all charged particles,
one can investigate the $p_t$ fluctuations of the negative and positive charges
independently as well as the cross correlation between them.
Let us define
\be
p_t^\pm = {P_t^\pm\over N_+}
=
{\sum_p n_p^\pm(p) p_t(p) \over \sum_p n_p^\pm(p)}
\ee
where $n_p^\pm(p)$ is the momentum distribution of 
the positively (negatively) charged particles.
Using Eq.(\ref{eq:ch4:ratio_fluct}), we then have
\be
\lefteqn{
\ave{\left(\delta p_t^\pm\right)^2} 
}
\non
& = &
\ave{p_t^\pm}^2
\sum_{p,q,\pm,\pm} 
\left(
{p_t(p) p_t(q)\over \ave{P_t^\pm}^2}
+
{1 \over \ave{N^\pm}^2}
-
{p_t(p) + p_t(q) \over \ave{N^\pm}\ave{P_t^\pm}}
\right)
\Delta_{p,q}^{\pm,\pm} 
\non
&=&
\frac{1}{\ave{N_\pm}^2} \sum_{p,q} 
\lb p_t(p) - \ave{{p_t^\pm}} \rb 
\lb p_t(q) - \ave{{p_t^\pm}} \rb
\Delta_{p,q}^{\pm,\pm}
\ee
where to get the last line we used
$\ave{p_t^\pm} = \ave{P_t^\pm}/\ave{N_\pm}$.
For the unlike-sign pairs, we use
Eq.(\ref{eq:unlike_corr}) to get
\be
\ave{\delta p_t^+ \delta p_t^-}
&=&
\frac{1}{\ave{N_+}\ave{N_-}} \sum_{p,q} 
\lb p_t(q) - \ave{p_t^+} \rb 
\lb p_t(p) - \ave{p_t^-} \rb 
\Delta_{p,q}^{+,-}
\ee
Therefore it is natural to define the measure of dynamic fluctuation as
\be
\Delta \sigma_{p_t^\pm}^2
\equiv
\ave{N_\pm} \ave{(\delta p^\pm_t)^2} - \incl{(\delta p^\pm_t)^2}
\ee
For unlike-sign pairs, $\incl{\delta p_t^+\delta p_t^-} = 0$
so that
\be
\Delta \sigma_{p_t^{+-}}^2
\equiv
\sqrt{\ave{N_+}\ave{N_-}} \ave{\delta p_t^+\delta p_t^-} 
\ee

This then allows us to define the
`charge-independent' (CI) and `charge-dependent'
(CD) combinations as used in the STAR and the PHENIX 
collaborations\cite{Ray:2002md}
\be
\ave{N}\Delta\sigma^2_{\rm CI, \, CD} (p_t)
& \equiv & 
\ave{N_+}\Delta \sigma_{p_t^+}^2
+
\ave{N_-}\Delta \sigma_{p_t^-}^2
\pm
2
\sqrt{\ave{N_+}\ave{N_-}} \Delta \sigma_{p_t^{+-}}^2
\non
\label{eq:ch4:pt_cd}
\ee
Here the $(+)$-sign in front of last  term leads to the
`charge-independent' and the $(-)$-sign to the `charge dependent' combinations,
respectively.   
Under the quite reasonable assumption that
\be
\frac{\ave{P_t^+}}{\ave{N_+}} = \frac{\ave{P_t^-}}{\ave{N_-}} 
= \frac{\ave{P_t^+ + P_t^-}}{\ave{N_+ + N_-}} = 
 \ave{p_t}
\ee
and 
$\incl{(\delta p_t^+)^2} = \incl{(\delta p_t^-)^2} =
\incl{(\delta p_t)^2}$, this can be rewritten as
\be
& & \lefteqn{\ave{N_{\rm ch}}\Delta\sigma^2_{\rm CI, \, CD} (p_t) =}
\non & & {} \qquad\qquad
\sum_{p,q} 
\lb p_t(p) - \ave{p_t} \rb 
\lb p_t(q) - \ave{p_t} \rb 
\lb \Delta_{p,q}^{+,+} + \Delta_{p,q}^{-,-} 
\pm 2 \Delta_{p,q}^{+,-} \rb
\non & & {} \qquad\qquad
-
2 \ave{N_{\rm ch}} \, \incl{(\delta p_t)^2}
\ee
where the assignment of plus and minus signs are as in Eq.\reff{eq:ch4:pt_cd}.

Notice the the combinations of the basic correlator $\Delta$ in the
`charge-dependent' combination is the same as that of the net 
charge fluctuation Eq.\reff{eq:ch4:delta_Q} . 
Likewise, the `charge-independent' combination resembles the
fluctuation of the fluctuations of the number of charged particles, eq. 
\reff{eq:ch4:delta_n_ch}. As we discussed in detail 
in section \ref{sec:corr},
charge conservation suppresses the fluctuations of the net-charge in a
subsystem.  Therefore, one would expect, that also the `charge-dependent'
transverse momentum fluctuations are suppressed, since they are nothing but a
different moment of the same correlator. This is seen in the preliminary data
of the STAR collaboration \cite{Ray:2002md}. 

Other correlation, such as resonances and flow may also affect the transverse
momentum fluctuations. In Ref.\cite{Stephanov:1999zu} 
these have been estimated
and found to be small, of the order of $1-2 \%$.

\subsection{Fluctuations of particle ratios}
The fluctuation of particle ratios are also presently studied. Consider the
ratio of particles ``$1$'' and ``$2$''
\be
R_{12} = \frac{N_1}{N_2}
\ee 
then with $A = N_1$ and $B=N_2$ the dispersion is given by
eq. \reff{eq:ch4:delta_r1} as
\be
\ave{\delta R_{12}^2} = \frac{\ave{N_1}^2}{\ave{N_2}^2} 
\lb 
\frac{(\delta N_1)^2}{\ave{N_1}^2} + \frac{(\delta N_2)^2}{\ave{N_2}^2} 
-2 \frac{\delta N_1 \, \delta N_2 }{\ave{N_1} \ave{N_2}}
\rb
\label{eq:ch4:part_ratio}
\ee
Notice that the last term introduces correlations between the particles which may
reduce the fluctuations. One example for these correlations are
hadronic resonances.  Consider, for example 
the ratio of positively and negatively charged pions $R = \frac{\pi^+}{\pi^-}$. 
The the presence of neutral rho-mesons, which decay into $\pi^+$ and $\pi^-$
reduce the fluctuations. This can be easily understood by considering a gas
made out of only neutral rho mesons. Independent, how large the fluctuations
of these are, in every event on has as many $\pi^+$ as $\pi^-$, and hence the
fluctuations of their ratio vanish. This effect can be utilized to estimate
the number of resonance in the system at chemical freeze-out
\cite{Jeon:1999gr}.

Next, consider the situation 
where one particle species is much more abundant then
the other, i.e. $\ave{N_2} \gg \ave{N_1}$. This is the case of the
fluctuations of the kaon to pion ratio, as investigated by the NA49
collaboration. Assuming that the number fluctuations are approximately
Poisson, then the second term in eq. \reff{eq:ch4:part_ratio} dominates
\be
\frac{(\delta N_1)^2}{\ave{N_1}^2} \simeq \frac{1}{\ave{N_1}} \gg
\frac{1}{\ave{N_2}} \simeq \frac{(\delta N_2)^2}{\ave{N_2}^2} 
\ee
so that
\be
\frac{\ave{\delta R_{12}^2}}{\ave{R_{12}}^2}
\simeq  \frac{1}{\ave{N_1}}
\left(1 + {\cal O}\lb {\ave{N_1}}/{\ave{N_2}}\rb \right)
\ee
In other words, if one particle species is much more abundant then the other,
the correlations among the particles have to be very strong to be visible in
experiment. To give some numbers, in a standard resonance gas
\cite{redlich_here} the correlations due to resonance lead to $30 \%$
corrections in case of the $\pi^+/\pi^-$ ratio and only to $4 \%$ in case of
the $K/\pi$ ratio \cite{Jeon:1999gr}.

Thus, ratios of equally abundant particles are best suited to expose
possible correlations. The simplest, and arguably most interesting is that of
positively over negatively charged particles
\be
R_{+-} &=& \frac{N_+}{N_-}
\\
\ave{\delta R_{+-}^2} &=& \ave{R_{+-}}^2 
\lb 
\frac{\ave{\delta N_+^2}}{\ave{N_+}^2} + \frac{\ave{\delta N_-^2}}{\ave{N_-}^2} 
-2 \frac{\ave{\delta N_+ \, \delta N_- }}{\ave{N_-} \ave{N_-}}
\rb
\ee
In the limit that the net charge $\ave{Q} = \ave{N_+ - N_-}$ is much smaller
than the number of charged particles $\ave{N_{\rm ch}} = \ave{N_+ + N_-}$,
or $\ave{Q} \ll \ave{N_{\rm ch}}$
\be
\ave{R_{+-}} \simeq 1; \,\,\, \ave{N_+} \simeq \ave{N_-} \simeq 
\frac{\ave{N_{\rm ch}}}{2}
\ee
so that
\be
\ave{\delta R_{+-}^2} 
&=& \frac{4}{\ave{N_{\rm ch}}^2} 
\ave{ \delta N_+^2 + \delta N_-^2 - 2 \delta N_+ \, \delta N_-}
\non
&=& 4 \frac{\ave{\delta Q^2}}{\ave{N_{\rm ch}}^2} 
\ee
Since the number of charged particles is a measure for the entropy of the
system, $\ave{N_{\rm ch}} \propto S$, the observable
\be
D \equiv \ave{N_{\rm ch}} \ave{\delta R_{+-}^2}  = 
4 \frac{\ave{\delta Q^2}}{\ave{N_{\rm ch}}} \propto
\frac{\ave{\delta Q^2}}{S}
\ee
is a measure for the charge fluctuations per entropy. And this, as discussed
in detail 
in section \ref{sec:dens_fluct}, is an observable for the existence of
the Quark Gluon Plasma. As discussed in detail in \cite{Jeon:2000wg}
for a pion gas, $D_{\rm pion-gas} = 4$ whereas for a QGP, 
$D_{\rm QGP} \simeq 1-1.5$, where the
uncertainty arises from relating the entropy $S$ with the number of charged
particles $\ave{N_{\rm ch}}$. 
Hadronic resonances introduce additional correlations,
which reduce the value of the pion gas to $D_{\rm hadron-gas} \simeq 3$, but
still a factor of 2 larger then the value for the QGP.

In principle, one could directly measure the ratio 
$\frac{\ave{\delta Q^2}}{\ave{N_{\rm ch}}}$, without going through ratio
  fluctuations. However, since the net charge  is an extensive quantity, this
  will introduce volume fluctuations into the measurement. Only in the limit
  that the total charge of the system is zero, volume fluctuations do not
  contribute to the order considered here \cite{Jeon:2001ue}.

The key question of course is, how can these reduced fluctuations be observed
 in the final state which consists of hadrons. Should one not expect that
 the fluctuations will be those of the hadron gas? 
 The reason, why it should be possible to see the charge fluctuations of the
 initial QGP has to do with the fact that charge is a conserved quantity. 
 Imagine one measures in each event the net charge in a given rapidity
 interval $\Delta y$ such that 
 \be
 \Delta y_{coll} \ll \Delta y \ll \Delta y_{max}
 \ee
 where $\Delta y_{max}$ is the width of the total charge distribution and 
$\Delta y_{coll}$ is the typical rapidity shift due to hadronization and
 re-scattering in the hadronic phase.
 If, as it is expected, strong longitudinal flow develops already in the
 QGP-phase, the number of charges inside the rapidity window $\Delta y$ 
 for a given event is essentially frozen in.
 And if $\Delta y \gg \Delta y_{coll}$ 
neither hadronization nor the subsequent collisions
 in the hadronic phase will be very effective to transport charges in and out
 of this rapidity window. Thus, the \ebe charge-fluctuations measured at the 
 end  reflect those of the initial state, when the longitudinal flow is
 developed. Ref. \cite{Shuryak:2000pd} arrives at the same conclusion  
 on the basis of a Fokker-Planck type equation describing the relaxation of
 the charge fluctuation in a thermal environment.

\begin{figure}[htb]
  \begin{center}
  \epsfxsize=\tw
  \epsffile{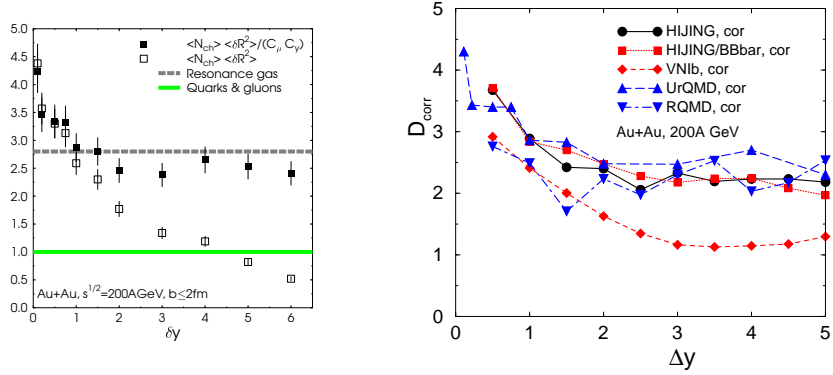}
 \end{center}
\vspace{-0.8cm}
  \caption{Left: 
Charge fluctuations for different rapidity windows as obtained
  in the URQMD model. Open symbols: without correction for global 
    charge conservation. Full symbols: With
    correction for global charge conservation. Right: Results for different
  transport model including the parton cascade
  \protect\cite{Zhang:2002dy}. These results are corrected for charge conservation.}
  \label{fig:ch4:bleicher}
\end{figure}

In Fig.\ref{fig:ch4:bleicher} we show the results of an URQMD calculation
\cite{Bleicher:2000ek} (left figure), 
where the variable $D$ is plotted versus the size of
the rapidity window $\Delta y$.
For large $\Delta y$ the results have to be corrected for charge conservation
effects; if all charges are accepted, global charge conservation leads to
vanishing fluctuations (open symbols in Fig.\ref{fig:ch4:bleicher}). 
This can be  corrected for as explained  in the previous section 
(for details see \cite{Bleicher:2000ek}). 
The resulting
values for $D$ are shown as full symbols in Fig.\ref{fig:ch4:bleicher}. They
agree nicely with the prediction for the resonance gas, as they should, since
the URQMD model does not contain any partonic degrees of freedom. For small
$\Delta y$ the correlations imposed by the resonances are lost, because 
only one of the decay products is accepted. As a result we see an increase 
of $D$. 
The right figure shows a comparison of several transport models, including the
Parton Cascade model \cite{Geiger:1992nj}. 
This model starts with partons in the
initial state, but has some model for hadronization included. If the general
ideas about the reduced charge fluctuations are correct, this model should
lead to smaller values of $D$, and it does.

For very small $\Delta y$, when $\ave{N} \simeq 1$, the ratio $D$ is
not well defined for events with $N_- = 0$, and, therefore, 
cannot serve as a observable. Alternative observables, measuring the same
quantity have been proposed and studied in \cite{Pruneau:2002yf}.

\subsection{Kaon Fluctuations}

\begin{figure}[htb]
  \epsfysize=7cm
  \centerline{\epsfbox{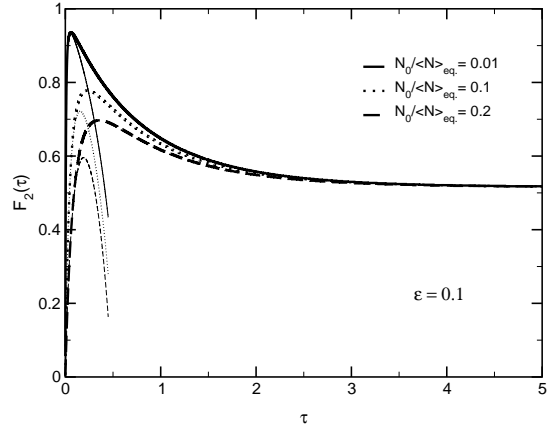}}
  \caption{Time evolution of $F_2$ for various initial kaon numbers. The thin
  lines are analytical results for early times.}
  \label{fig:ch4:kaon_fluct}
\end{figure}

As discussed in section \ref{sub:fluct_can} at low energy the strangeness
conservation has to be treated explicitly. This can be utilized to
determine the degree of equilibration reached in these collisions
\cite{Jeon:2001ka}. By measuring the ratio 
\be 
F_2 = \frac{ \ave{N_K (N_K -1)}}{\ave{N_K}^2}
\ee 
of the number of  kaon pairs over the square of the inclusive number of kaons
a factor two sensitivity on the degree of equilibrium can be obtained. 
From transport models (see e.g. \cite{Bratkovskaya:1997pj}) 
it is known, that most of the kaons are produced in
secondary collisions, i.e. during the evolution of the fireball. Transport
calculations also find that the equilibration time of kaons is of the order of
$500 \rm fm/c$, which is about ten time the lifetime of the system. Hence
these models would not predict equilibration of the kaons. However, observed
particle ratios {\em including } kaons 
are also consistent with a thermal model description (for a review see 
\cite{redlich_here}). The measurement of $F_2$ can resolve this issue and point
to new physics such as multi-particle processes, in medium effects etc., 
if indeed it is found to be consistent with equilibrium.
This is
demonstrated in Fig.\ref{fig:ch4:kaon_fluct} where
the time evolution of $F_2$ for several
initial kaon numbers is shown. 
In all cases, $F_2$ quickly rises close to $F_2 \simeq 1$
before it settles at the final equilibrium value of $F_2^{\rm eq.} \simeq 1/2$.
Thus, by measuring $F_2$ one can directly determine how close to chemical
equilibrium the system has developed, before it freezes out.
If the predictions of the transport models are correct, then a value of $F_2
\simeq 1$ should be found. If, on the other hand, equilibrium is indeed reached
in these collisions, then $F_2 \simeq \frac{1}{2}$. 
In principle a similar  measurement can also be done at higher energies for
charmed mesons. To which extent this is technically feasible is another 
question.

Let us conclude the section by mentioning other observables. In the context of
so called DCC production \cite{Rajagopal:1995bc}, the fluctuations of the
fraction of neutral pions is considered a useful signal. Also the fluctuations
of the elliptic flow has be proposed as an useful observable
\cite{Mrowczynski:2002bw}, which may reveal new configurations such as
sphalerons, which could be created in heavy ion collisions.


%% file: chapter_5.tex
\section{Experimental Situation}
\label{sec:experiment}

   \subsection{Fluctuation in elementary collisions}
   \label{sec:pp-stuff}

    By now, it is clear that Quantum Chromodynamics is the right theory of
    strong interaction.  However, essential non-perturbative
    nature of the strong interaction still remains a difficulty.  In heavy ion
    collisions, this is more manifest in the sense that even the hard
    part of the spectrum is strongly influenced by the surrounding soft medium
    through multiple scatterings.  Therefore, before we begin to consider the
    experimental results from heavy ion collisions, 
    it is crucial that we understand elementary collision results such as
    proton-proton collision results since these elementary collision results
    can provide unbiased baseline.

    A large amount of data for $pp$ collisions at various beam energies
    were taken from Fermi Lab in the 1970's
    \cite{Bromberg:1974ga,Kafka:1977py,Kafka:1975cz,Kafka:1979pp} (see also
    Ref.\cite{Whitmore:1976ip}). 
    Among them, data at the beam energy of $205\,\GeV$ were most thoroughly
    analyzed by Kafka et.al.\cite{Kafka:1975cz,Kafka:1977py}. 
    First, let us consider the `charge fluctuation study'. 
    In the case of $pp$ collisions, the definition of `charge fluctuation', 
    however, differs from ours.  Let us define the charge transfer
    \be
    u(y) = {1\over 2}\left[ Q_f(y) - Q_b(y) \right]
    \ee
    where $Q_f(y)$ is the sum of the charges of the
    particles with rapidity larger than $y$ and 
    $Q_b(y)$ is the sum of the charges of the 
    particles with rapidity smaller than than $y$.
    We then define the charge transfer fluctuation
    \be
    \ebeave{\delta u(y)^2} = \ebeave{u^2(y)} - \ebeave{u(y)}^2
    \ee
    This quantity is usually referred to as `charge fluctuation' in literature
    dealing with proton-proton collisions.
    This charge transfer fluctuation 
    is certainly not the same as what we have been discussing so far which
    is the charge fluctuation within a given phase space window $\Delta\eta$.
    However, as we will show shortly, $\ave{\delta u(y)^2}$
    is intimately related to
    the charge correlation functions. By considering 
    the charge transfer fluctuation, we can then put a constraint on possible
    forms of the correlation functions.

    To do so, consider again our simple model defined by Eq.(\ref{eq:mom}).  
    As before, we impose the conditions
    $ M_+ - M_- = Q_c = 0$, $f_+ = f_- = f$ and $h_+ = h_- = h$. 
    After straightforward but tedious algebra, we obtain 
    \be
    \ebeave{\delta u(y)^2}
    & = &
    \left(\ave{M_+} + \ave{M_-}\right)
    \int_y^{\infty} dy'\, f(y') \int^y_{-\infty} dy''\, f(y'')
    \non
    & & {}
    +
    2\ave{M_0} 
    \int_y^{\infty} dy_+ \int_{-\infty}^y dy_- f_0(y_+,y_-)
    \ee
    using the rapidity $y$ as the phase space variable.

    In Ref.\cite{Kafka:1975cz}, it is shown that the data satisfies
    \be
    \ebeave{\delta u(y)^2} \approx 0.62 {dN_{\rm ch}\over dy}
    \label{eq:kafka_dndy}
    \ee
    at the beam energy of
    $205\,\GeV$ and the shape of $dN_{\rm ch}/dy$ is very well represented by 
    a Gaussian with $\sigma^2_y \approx \log(\sqrt{s}/2m_N) \approx 2.2$.
    This result puts a condition on possible forms of $f$ and $f_0$.
    In our model, the rapidity distribution of the charged particle is 
    \be
    {dN_{\rm ch}\over dy}
    =
    \left(\ave{M_+} + \ave{M_-}\right) f(y)
    +
    2 \ave{M_0} h(y)
    \ee
    First, 
    let us see if charge particles alone can satisfy 
    Eq.(\ref{eq:kafka_dndy}).
    For this to be possible we must have 
    $\int_y^{\infty} dy'\, f(y') \int^y_{-\infty} dy''\, f(y'') \propto f(y)$
    and $\int f = 1$.
    These conditions are satisfied 
    by
    \be
    f(y) = {1\over 2 \tilde{y}}\, {\rm sech}^2(y/\tilde{y})
    \ee
    where $\tilde{y}$ is a constant specifying the width of the
    distribution.\footnote{This can be easily verified using 
    \be
    {d\tanh(x)\over dx} 
    = 1 - \tanh^2(x) = {\rm sech}^2(x)
    \ee}
    This form of $f(y)$ is however, too sharply peaked to be
    consistent with the data.  On the other hand a Gaussian with the above
    $\sigma_y^2$ can {\em approximately} satisfy the condition
    $\int_y^{\infty} dy'\, f(y') \int^y_{-\infty} dy''\, f(y'') \propto f(y)$.
    But in this case, the proportionality constant at $y=0$ is close to 1.
    All this indicates that we need the correlation term involving 
    $f_0(y_+, y_-)$.

    The condition (\ref{eq:kafka_dndy}) can be approximately satisfied if
    $f_0$ has the following form
    \be
    f_0(y_1, y_2) = q(y_1 - y_2) g( (y_1+y_2)/2 )
    \ee
    where $q(y)$ is a sharply peaked function at $y=0$
    with a small width $\Delta$, and $g(y)\approx ({1/\ave{N}}) dN/dy$.
    Then it can be also shown that  $h(y) \approx g(y)$ and
    \be
    \int_y^\infty dy' \int^y_{-\infty} dy'' f_0(y', y'')
    \approx
    C\, \Delta \, h(y)
    \ee
    with some constant $C < 1$.
    In fact the authors of Ref.\cite{Kafka:1975cz}
    argued that their proton-proton result is consistent with
    having only the neutral clusters.  In our language, that corresponds to 
    $\ave{M_\pm} = 0$.  

    What the above consideration implies for our charge fluctuation is somewhat
    unclear.  To the authors' knowledge, direct measurement of charge
    fluctuation in the sense defined in the previous sections has not been
    carried out in proton-proton experiments.  What have been measured are the 
    $C_{\alpha\beta}$ functions defined in Eq.(\ref{eq:3_corr}).
    In Ref\cite{Kafka:1977py},
    the maximum height at $y_1 = y_2 = 0$ are given as
    \be
    & C_{++}(0,0) \approx 0.36 &
    \non
    & C_{--}(0,0) \approx 0.25&
    \non
    & C_{+-}(0,0) \approx 0.5 &
    \ee
    Using also the overall shape given in the same reference and the 
    fact\cite{Kafka:1975cz} that
    $\ave{N_{\rm ch}} \approx 6$ at this energy one can then 
    infer that  
    \be
    {\ave{\delta Q^2}_{\Delta\eta}\over \ave{N_{\rm ch}}_{\Delta\eta}}
    \approx 0.5 \sim 0.6
    \ee
    within $-1 < y < 1$ {\rm without} any charge conservation corrections.
    Correcting for charge conservation is a difficult task to perform. 
    If as asserted in Ref.\cite{Kafka:1975cz} all particles are produced via
    neutral clusters, then there is no correction to perform.  
    If however, some charged particles are emitted independently, then this is
    the lower bound.
    A more thorough analysis of the available data can undoubtedly yield more
    accurate estimate of the charge fluctuation.  However, that is clearly
    beyond the scope of this review.  We just make a remark here that our
    estimate of the charge fluctuation in QGP scenario  
    ${\ave{\delta Q^2}_{\Delta\eta}/\ave{N_{\rm ch}}_{\Delta\eta}}
    \approx 0.25 \sim 0.3$, 
    seems to be still substantially 
    smaller than the above proton-proton collision result.

\subsection{Fluctuations in Heavy Ion Reactions}
\label{sec:HI_experiment}

\begin{figure}[htb]
  \epsfxsize=\tw
  \centerline{\epsfbox{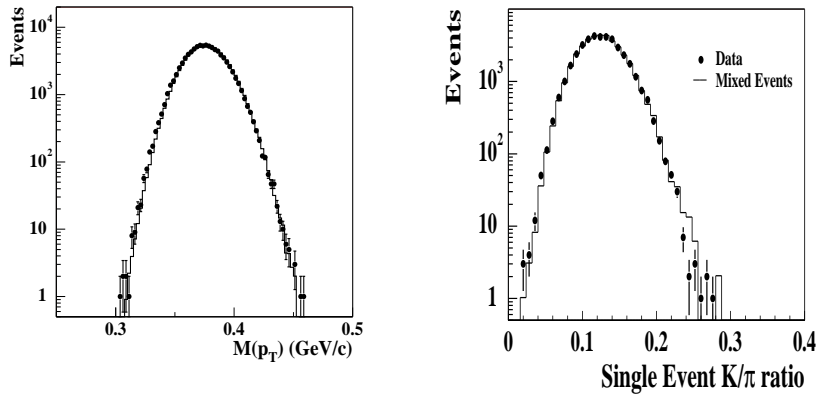}}
  \caption{Results for the fluctuations of the mean transverse momentum (left)
  and kaon to pion ratio (right). Both results are from the NA49 collaboration.
  }
  \label{fig:ch5:na49a}
\end{figure}

Unfortunately at the time of the writing of this review, very few published
data on fluctuations in heavy ion collisions are available. Quite a few
preliminary results are being discussed at conferences, which we will briefly
mention. But we feel that an in depth discussion of these results prior to
publication is not appropriate.   
The pioneering event-by-event studies have  been carried out by the NA49 
collaboration. They have analyzed the fluctuations of the mean transverse
momentum \cite{Appelshauser:1999ft} and the kaon to pion ratio 
\cite{Afanasev:2000fu}. Both measurements have been carried out at at the CERN
SPS at slightly forward rapidities. 
In Fig. \ref{fig:ch5:na49a} the resulting distributions
are shown together with that from mixed events (histograms). In both cases the
mixed event can essentially account for the observed signal, leaving little
room for genuine dynamical fluctuations. Specifically, NA49 gives 
$\Phi_{p_t} = 0.6 \pm 1.0 \, \rm MeV$\footnote{For the definition of
  $\Phi_{p_t}$ see previous section.} for the transverse momentum fluctuations,
which is compatible with zero. For the kaon to pion ratio they extract a width
due to non-statistical fluctuations of $\sigma_{non_stat} =2.8 \% \pm 0.5
\%$, which would be compatible form the expectations of resonance decays
\cite{Jeon:1999gr}. 
  
The PHENIX collaboration recently also reported their results on mean 
transverse momentum and energy fluctuations at RHIC energies 
\cite{Adcox:2002pa}. 
Similar to NA49, their result is
compatible with the statistical fluctuations only, leaving no room for
significant dynamical fluctuations. Their measurement was around mid-rapidity
with a small azimuthal acceptance of $\Delta \Phi = 58^\circ$

In contrast to that preliminary results by both the CERES  collaboration 
\cite{ceres_private} at the CERN SPS, as well as the STAR
collaboration \cite{Ray:2002md} at RHIC report significant dynamical 
fluctuations of
the mean transverse momentum. Both measure at mid-rapidity. It remains to be
seen if these difference in measurements can be attributed simply to the
different acceptance regions covered.

As far as the charge fluctuations are concerned, first results 
have been reported. PHENIX \cite{Adcox:2002mm} at RHIC
which measures with a small rapidity acceptance, finds charge fluctuations
consistent with a resonance gas. CERES \cite{ceres_private} and NA49
\cite{Blume:2002mr}, which both measure at SPS energies, report preliminary
results on charge fluctuations, which are consistent with a pure pion gas.   
However, at the SPS the overall rapidity distribution is rather narrow, so
that the correlation effect of the resonance gets lost when correcting for
charge conservation \cite{Zaranek:2001di}.  But certainly, non of the
measurements is even close to the prediction for the QGP. 
These findings
have prompted ideas, that possibly a constituent quark plasma, without gluons,
has been produced \cite{Bialas:2002hm}. However, the measurement of additional
observables would be needed in order to distinguish this from a hadronic gas. 

But maybe the present range of $\Delta y$ is so small, that the charge
fluctuations have time to assume the value of the resonance gas. As shown in
Fig. \ref{fig:ch4:bleicher}, the results for the parton cascade arrive at the
predicted value for the QGP only for $\Delta y \geq 3$. None of the present
experiments has such a coverage yet and thus a 
detailed analysis of $D$ as a function of $\Delta y$ is needed, before any
firm conclusions can be drawn.

The Balance function measurement at $\sqrt{s} = 130\,\GeV$ has been reported
by STAR collaboration\cite{StarBalance}.  Going from peripheral to central
collisions, the width of balance function steadily decreases.  
The trend is what one would expect if more of the system is filled with a
QGP as the collision becomes more central.  However,
since the reduction is only about 20\,\% going from most perpheral to most
central, it is not yet clear whether this signals the presence of a QGP or
more mundane effect such as the strong flow.


%% file: conclusions.tex
\section{Conclusions and Outlook}
\label{sec:conclusions}
In this review we discussed several aspects of the emerging field of
event-by-event fluctuations in heavy ion collisions. We have briefly reviewed 
the basics concepts of fluctuations in a thermal system, and we have discussed
the possible connection between a idealized thermal system and a ``real world''
heavy ion collision. Furthermore we have introduced the variables used in
order to present fluctuations and discussed the relation between them. Finally
we have present the presently published data, which are rather limited. But
this is soon going to change.

Crucial for the understanding of fluctuations is the connections to correlation
functions. While the concept of fluctuations might be more intuitive and thus
might lead us to new insights and approaches, the
relevant information is encoded in the correlation functions. 
For example, consider the
charge fluctuations, which we have discussed in some detail.
It is true that like-sign and unlike-sign correlation functions 
encode all the necessary information on the charge fluctuations.  
However the physical meanings of those encoded information 
can be made clear only through considering charge fluctuations in and out of
QGP and its connection to the relevant features of the correlation functions.
But certainly from the observational point of view, the information to be
extracted are the correlation functions between all possible quanta. Once they
are available, the relevant observables can be readily obtained by folding the
correlation functions with the appropriate variables

On important aspect relevant to heavy ion collisions is the finiteness of the
system. Thus approximations which are  valid for a large system, such as the
grand-canonical ensemble. Appropriate corrections have to be applied to either
theory or data in order to extract the properties of the bulk matter. While
this is not new to heavy ion physics, for the case of fluctuations these
methods still require more refinement.

Since the study of
fluctuations is a new and developing area of heavy ion physics, this review
can only be some kind of snapshot at a ``hopefully'' fast developing field.
By no means can it be comprehensive. Most of the experimental data are
available, if at all, only in preliminary form. Furthermore analysis methods
and possible new observable based on fluctuations are being developed as we
write the review.  
Therefore our idea was to present the basic concepts as well as 
some of the necessary formalism in a consistent and understandable way. To
which extent we have succeeded is beyond our judgment.

\vfill

\noindent
{\bf Acknowledgments}

\noindent
The authors thank S.~Voloshin, L.~Ray, C.~Gale, C.~Pruneau and S.~Gavin 
for suggestions and discussions.
V.K.~was supported by the Director, 
Office of Science, Office of High Energy and Nuclear Physics, 
Division of Nuclear Physics, and by the Office of Basic Energy
Sciences, Division of Nuclear Sciences, of the U.S. Department of Energy 
under Contract No. DE-AC03-76SF00098.
S.J.~thanks RIKEN
BNL Center and U.S. Department of Energy [DE-AC02-98CH10886] for
providing facilities essential for the completion of this work.
S.J.~is also supported in part by the Natural Sciences and
Engineering Research Council of Canada. 

%% file: ebe_review.bbl
\begin{thebibliography}{10}

\bibitem{Cobe}
"e.g. COBE web page at"
  "http://space.gsfc.nasa.gov/astro/cobe/cobe\_home.html".

\bibitem{Adcox:2002pa}
K.~Adcox et~al.
\newblock Event-by-event fluctuations in mean p(t) and mean e(t) in
  s(nn)**(1/2) = 130-gev au + au collisions.
\newblock {\em Phys. Rev.}, C66:024901, 2002.

\bibitem{Adcox:2002mm}
K.~Adcox et~al.
\newblock Net charge fluctuations in au + au interactions at s(nn)**(1/2) =
  130-gev.
\newblock {\em Phys. Rev. Lett.}, 89:082301, 2002.

\bibitem{Afanasev:2000fu}
S.~V. Afanasev et~al.
\newblock Event-by-event fluctuations of the kaon to pion ratio in central pb +
  pb collisions at 158-gev per nucleon.
\newblock {\em Phys. Rev. Lett.}, 86:1965--1969, 2001.

\bibitem{Appelshauser:1999ft}
H.~Appelshauser et~al.
\newblock Event-by-event fluctuations of average transverse momentum in central
  pb + pb collisions at 158-gev per nucleon.
\newblock {\em Phys. Lett.}, B459:679--686, 1999.

\bibitem{Asakawa:2000wh}
Masayuki Asakawa, Ulrich~W. Heinz, and Berndt Muller.
\newblock Fluctuation probes of quark deconfinement.
\newblock {\em Phys. Rev. Lett.}, 85:2072--2075, 2000.

\bibitem{Baym:1995cz}
Gordon Baym, B.~Blattel, L.~L. Frankfurt, H.~Heiselberg, and M.~Strikman.
\newblock Correlations and fluctuations in high-energy nuclear collisions.
\newblock {\em Phys. Rev.}, C52:1604--1617, 1995.

\bibitem{Baym:1999up}
Gordon Baym and Henning Heiselberg.
\newblock Event-by-event fluctuations in ultrarelativistic heavy-ion
  collisions.
\newblock {\em Phys. Lett.}, B469:7--11, 1999.

\bibitem{Berdnikov:1999ph}
Boris Berdnikov and Krishna Rajagopal.
\newblock Slowing out of equilibrium near the qcd critical point.
\newblock {\em Phys. Rev.}, D61:105017, 2000.

\bibitem{Bertsch:1994qc}
George~F. Bertsch.
\newblock Meson phase space density from interferometry.
\newblock {\em Phys. Rev. Lett.}, 72:2349--2350, 1994.

\bibitem{Bialas:2002hm}
A.~Bialas.
\newblock Charge fluctuations in a quark antiquark system.
\newblock {\em Phys. Lett.}, B532:249--251, 2002.

\bibitem{Bialas:1999tv}
A.~Bialas and V.~Koch.
\newblock Event by event fluctuations and inclusive distributions.
\newblock {\em Phys. Lett.}, B456:1--4, 1999.

\bibitem{Blaizot:2002xz}
J.~P. Blaizot, E.~Iancu, and A.~Rebhan.
\newblock Comparing different hard-thermal-loop approaches to quark number
  susceptibilities.
\newblock 2002.

\bibitem{Blaizot:2001vr}
J.~P. Blaizot, Edmond Iancu, and A.~Rebhan.
\newblock Quark number susceptibilities from htl-resummed thermodynamics.
\newblock {\em Phys. Lett.}, B523:143--150, 2001.

\bibitem{Bleicher:2000ek}
M.~Bleicher, S.~Jeon, and V.~Koch.
\newblock Event-by-event fluctuations of the charged particle ratio from
  non-equilibrium transport theory.
\newblock {\em Phys. Rev.}, C62:061902, 2000.

\bibitem{Blume:2002mr}
C.~Blume et~al.
\newblock Results on correlations and fluctuations from na49.
\newblock 2002.

\bibitem{Bratkovskaya:1997pj}
E.~L. Bratkovskaya, W.~Cassing, and U.~Mosel.
\newblock Analysis of kaon production at sis energies.
\newblock {\em Nucl. Phys.}, A622:593--604, 1997.

\bibitem{redlich_here}
P~Braun-Munzinger, K~Redlich, and J~Stachel.
\newblock {\em this review volume}.

\bibitem{Bromberg:1974ga}
C.~M. Bromberg et~al.
\newblock Pion production in p p collisions at 102-gev/c.
\newblock {\em Phys. Rev.}, D9:1864--1871, 1974.

\bibitem{Cleymans:1991mn}
J.~Cleymans, K.~Redlich, and E.~Suhonen.
\newblock Canonical description of strangeness conservation and particle
  production.
\newblock {\em Z. Phys.}, C51:137--141, 1991.

\bibitem{ceres_private}
CERES collaboration.
\newblock {\em to be published}, 2002.
\newblock prelimnary results, private communication.

\bibitem{Doring:2002qa}
M.~Doering and V.~Koch.
\newblock Event-by-event fluctuations in heavy ion collisions.
\newblock {\em Acta Phys. Polon.}, B33:1495--1504, 2002.

\bibitem{Eletsky:1993hv}
V.~L. Eletsky, J.~I. Kapusta, and R.~Venugopalan.
\newblock Screening mass from chiral perturbation theory, virial expansion and
  the lattice.
\newblock {\em Phys. Rev.}, D48:4398--4407, 1993.

\bibitem{StarBalance}
J.Adams et.al.
\newblock Narrowing of the balance function with centrality in au + au
  collisions s(nn)**(1/2) = 130-gev.
\newblock 2003.

\bibitem{Fodor:2001pe}
Z.~Fodor and S.~D. Katz.
\newblock Lattice determination of the critical point of qcd at finite t and
  mu.
\newblock {\em JHEP}, 03:014, 2002.

\bibitem{Gavai:2002kq}
Rajiv~V. Gavai and Sourendu Gupta.
\newblock The continuum limit of quark number susceptibilities.
\newblock {\em Phys. Rev.}, D65:094515, 2002.

\bibitem{Gazdzicki:1992ri}
M.~Gazdzicki and S.~Mrowczynski.
\newblock A method to study 'equilibration' in nucleus-nucleus collisions.
\newblock {\em Z. Phys.}, C54:127--132, 1992.

\bibitem{Gazdzicki:1997gm}
Marek Gazdzicki.
\newblock A method to study chemical equilibration in nucleus nucleus
  collisions.
\newblock {\em Eur. Phys. J.}, C8:131--133, 1999.

\bibitem{Geiger:1992nj}
Klaus Geiger and Berndt Muller.
\newblock Dynamics of parton cascades in highly relativistic nuclear
  collisions.
\newblock {\em Nucl. Phys.}, B369:600--654, 1992.

\bibitem{Gottlieb:1997ae}
S.~Gottlieb et~al.
\newblock Thermodynamics of lattice qcd with two light quark flavours on a
  16**3 x 8 lattice. ii.
\newblock {\em Phys. Rev.}, D55:6852--6860, 1997.

\bibitem{Heinz:1999rw}
Ulrich~W. Heinz and Barbara~V. Jacak.
\newblock Two-particle correlations in relativistic heavy-ion collisions.
\newblock {\em Ann. Rev. Nucl. Part. Sci.}, 49:529--579, 1999.

\bibitem{Heiselberg:2000ti}
H.~Heiselberg and A.~D. Jackson.
\newblock Anomalous multiplicity fluctuations from phase transitions in heavy
  ion collisions.
\newblock {\em Phys. Rev.}, C63:064904, 2001.

\bibitem{Heiselberg:2000fk}
Henning Heiselberg.
\newblock Event-by-event physics in relativistic heavy-ion collisions.
\newblock {\em Phys. Rept.}, 351:161--194, 2001.

\bibitem{Jeon:1999gr}
S.~Jeon and V.~Koch.
\newblock Fluctuations of particle ratios and the abundance of hadronic
  resonances.
\newblock {\em Phys. Rev. Lett.}, 83:5435--5438, 1999.

\bibitem{Jeon:2000wg}
S.~Jeon and V.~Koch.
\newblock Charged particle ratio fluctuation as a signal for qgp.
\newblock {\em Phys. Rev. Lett.}, 85:2076--2079, 2000.

\bibitem{Jeon:2001ka}
S.~Jeon, V.~Koch, K.~Redlich, and X.~N. Wang.
\newblock Fluctuations of rare particles as a measure of chemical
  equilibration.
\newblock {\em Nucl. Phys.}, A697:546--562, 2002.

\bibitem{Jeon:2001ue}
Sang-yong Jeon and Scott Pratt.
\newblock Balance functions, correlations, charge fluctuations and
  interferometry.
\newblock {\em Phys. Rev.}, C65:044902, 2002.

\bibitem{Kafka:1975cz}
T.~Kafka et~al.
\newblock Charge and multiplicity fluctuations in 205-gev/c p p interactions.
\newblock {\em Phys. Rev. Lett.}, 34:687--690, 1975.

\bibitem{Kafka:1977py}
T.~Kafka et~al.
\newblock One, two, and three particle distributions in p p collisions at
  205-gev/c.
\newblock {\em Phys. Rev.}, D16:1261, 1977.

\bibitem{Kafka:1979pp}
T.~Kafka et~al.
\newblock Correlations between neutral and charged pions produced in 300-gev/c
  p p collisions.
\newblock {\em Phys. Rev.}, D19:76, 1979.

\bibitem{Kapusta_book}
J.~Kapusta.
\newblock {\em Finite Temperature Field Theory}.
\newblock Cambridge University Press, 1989.

\bibitem{Ko:2000vp}
C.~M. Ko et~al.
\newblock Kinetic equation with exact charge conservation.
\newblock {\em Phys. Rev. Lett.}, 86:5438--5441, 2001.

\bibitem{Koch:2001zn}
V.~Koch, M.~Bleicher, and S.~Jeon.
\newblock Event-by-event fluctuations and the qgp.
\newblock {\em Nucl. Phys.}, A698:261--268, 2002.

\bibitem{Landau_Stat}
L~Landau and L.M. Lifshitz.
\newblock {\em Statistical Physics}.
\newblock Pergamon Press, New York, 1980.

\bibitem{Lin:2001mq}
Zi-wei Lin and C.~M. Ko.
\newblock Baryon number fluctuation and the quark gluon plasma.
\newblock {\em Phys. Rev.}, C64:041901, 2001.

\bibitem{Mrowczynski:1998vt}
Stanislaw Mrowczynski.
\newblock Transverse momentum and energy correlations in the equilibrium system
  from high-energy nuclear collisions.
\newblock {\em Phys. Lett.}, B439:6--11, 1998.

\bibitem{Mrowczynski:2002bw}
Stanislaw Mrowczynski and Edward~V. Shuryak.
\newblock Elliptic flow fluctuations.
\newblock 2002.

\bibitem{Prakash:2001xm}
Madappa Prakash, Ralf Rapp, Jochen Wambach, and Ismail Zahed.
\newblock Isospin fluctuations in qcd and relativistic heavy-ion collisions.
\newblock {\em Phys. Rev.}, C65:034906, 2002.

\bibitem{Pruneau:2002yf}
C.~Pruneau, S.~Gavin, and S.~Voloshin.
\newblock Methods for the study of particle production fluctuations.
\newblock 2002.

\bibitem{Rajagopal:1995bc}
Krishna Rajagopal.
\newblock The chiral phase transition in qcd: Critical phenomena and long
  wavelength pion oscillations.
\newblock 1995.

\bibitem{Ray:2002md}
R.~L. Ray.
\newblock Correlations, fluctuations, and flow measurements from the star
  experiment.
\newblock 2002.
\newblock Proceedings Quark Matter 2002, Nantes, France, July 2002.

\bibitem{reif_book}
F.~Reif.
\newblock {\em Fundamentals of Statistical and Thermal Physics}.
\newblock McGraw-Hill, 1984.

\bibitem{Shuryak:1998yj}
Edward~V. Shuryak.
\newblock Event-per-event analysis of heavy ion collisions and thermodynamical
  fluctuations.
\newblock {\em Phys. Lett.}, B423:9--14, 1998.

\bibitem{Shuryak:2000pd}
Edward~V. Shuryak and Misha~A. Stephanov.
\newblock When can long range charge fluctuations serve as a qgp signal?
\newblock {\em Phys. Rev.}, C63:064903, 2001.

\bibitem{Stephanov:1998dy}
Misha~A. Stephanov, K.~Rajagopal, and Edward~V. Shuryak.
\newblock Signatures of the tricritical point in {QCD}.
\newblock {\em Phys. Rev. Lett.}, 81:4816--4819, 1998.

\bibitem{Stephanov:1999zu}
Misha~A. Stephanov, K.~Rajagopal, and Edward~V. Shuryak.
\newblock Event-by-event fluctuations in heavy ion collisions and the {QCD}
  critical point.
\newblock {\em Phys. Rev.}, D60:114028, 1999.

\bibitem{Stodolsky:1995ds}
L.~Stodolsky.
\newblock Temperature fluctuations in multiparticle production.
\newblock {\em Phys. Rev. Lett.}, 75:1044--1045, 1995.

\bibitem{Sturm:2002xq}
C.~Sturm et~al.
\newblock Kaon and antikaon production in dense nuclear matter.
\newblock {\em J. Phys.}, G28:1895--1902, 2002.

\bibitem{Tannenbaum:2001gs}
M.~J. Tannenbaum.
\newblock The distribution function of the event-by-event average p(t) for
  statistically independent emission.
\newblock {\em Phys. Lett.}, B498:29--34, 2001.

\bibitem{Voloshin:1999a}
S~Voloshin, V~Koch, and H.G. Ritter.
\newblock Event-by-event fluctuations in collective quantities.
\newblock {\em Phys. Rev.}, C60:0224901, 1999.

\bibitem{Whitmore:1976ip}
J.~Whitmore.
\newblock Multiparticle production in the fermilab bubble chambers.
\newblock {\em Phys. Rept.}, 27:187--273, 1976.

\bibitem{Zaranek:2001di}
Jacek Zaranek.
\newblock Measures of charge fluctuations in nuclear collisions.
\newblock {\em Phys. Rev.}, C66:024905, 2002.

\bibitem{Zhang:2002dy}
Q.~H. Zhang, V.~Topor~Pop, S.~Jeon, and C.~Gale.
\newblock Charged particle ratio fluctuations and microscopic models of nuclear
  collisions.
\newblock {\em Phys. Rev.}, C66:014909, 2002.

\end{thebibliography}
